\documentclass[fleqn,usenatbib]{mnras}

\usepackage[T1]{fontenc}

\DeclareRobustCommand{\VAN}[3]{#2}
\let\VANthebibliography\thebibliography
\def\thebibliography{\DeclareRobustCommand{\VAN}[3]{##3}\VANthebibliography}

\usepackage{graphicx}	
\graphicspath{ {figures//} }
\usepackage[utf8]{inputenc}		
\usepackage{pifont}
\usepackage{romannum}
\usepackage{multirow}
\usepackage[inline]{enumitem}
\usepackage{threeparttable}

\usepackage[dvipsnames]{xcolor}
\usepackage{hyperref}

\usepackage{listings} 
\usepackage{array} 
\usepackage{xifthen}
\usepackage{amsmath,amssymb,amsfonts}

\newcolumntype{'}{!{\vrule width 2pt}}

\newcolumntype{L}[1]{>{\raggedright\let\newline\\\arraybackslash\hspace{0pt}}m{#1}}
\newcolumntype{C}[1]{>{\centering\let\newline\\\arraybackslash\hspace{0pt}}m{#1}}
\newcolumntype{R}[1]{>{\raggedleft\let\newline\\\arraybackslash\hspace{0pt}}m{#1}}

\newcommand{\spm}[2]{\ensuremath{^{+#1}_{-#2}}}

\newcommand{\noop}[1]{}

\newcommand{\logM}[1][]{\ifthenelse{\isempty{#1}}{\log_{10}(M/\si{\Msun})}{\log_{10}(M_{#1}/\si{\Msun})}}
\newcommand{\ml}[1][]{\ifthenelse{\isempty{#1}}{M/L}{M_{#1}/L}}
\newcommand{\mlStar}{\ifmmode{\ml[{\star}]}\else{\(\ml[{\star}]\)}\fi}
\newcommand{\Rein}{\ifmmode{R_{\rm Ein}}\else{\(R_{\rm Ein}\)}\fi}
{}
{}
{}
{}
\newcommand\ft{\texttt{FORTRAN}}{}

\definecolor{maq}{HTML}{9F009F}
\definecolor{plum}{HTML}{581845}
\definecolor{pblue}{HTML}{0099F7}
\definecolor{pgreen}{HTML}{77DD77}
\definecolor{bbrown}{HTML}{BF9E86}
\definecolor{bgreen}{HTML}{C9DCAF}
\definecolor{dbrown}{HTML}{784128}
\definecolor{lime}{HTML}{c0ca33}

\newcommand{\tfo}[1]{\texttt{#1}}
\newcommand{\tso}[1]{\textsc{#1}}
\newcommand{\shw}{Schwarzschild}
\newcommand{\SZ}{\ifmmode{{\rm S}0}\else{\({\rm S}0\)}\fi}

\newcommand{\atomic}[2]{{\rm \nuclide{#1}\textsc{#2}}}

\newcommand{\ND}[1]{\(#1{\rm D}\)}
\newcommand{\lcdm}{\ifmmode{\Lambda{\rm CDM}}\else{\(\Lambda{\rm CDM}\)}\fi}
\newcommand{\eff}{\mathrm{e}}

\newcommand{\mfive}{\ifmmode{m_{0.5}}\else{\(m_{0.5}\)}\fi}
\newcommand{\mone}{\ifmmode{m_{1.0}}\else{\(m_{1.0}\)}\fi}

\newcommand{\chemFeH}   {\nuclide{[Fe/H]}}
\newcommand{\chemMgH}   {\nuclide{[Mg/H]}}
\newcommand{\chemCH}   {\nuclide{[C/H]}}
\newcommand{\chemNH}   {\nuclide{[N/H]}}
\newcommand{\chemNaH}   {\nuclide{[Na/H]}}
\newcommand{\chemCaH}   {\nuclide{[Ca/H]}}
\newcommand{\chemTiH}   {\nuclide{[Ti/H]}}

\newcommand{\chemZH}    {\nuclide{[Z/H]}}
\newcommand{\chemOH}    {\nuclide{[O/H]}}

\newcommand{\chemAlphFe}{\nuclide{[}{\ifmmode{\relax\alpha}\else{\(\relax\alpha\)}\fi}\nuclide{/Fe]}}
\newcommand{\snl}[1]{SNL\textendash {#1}}
\usepackage{tensor}
\let\scshape\relax 
\DeclareRobustCommand\scshape{%
  \not@math@alphabet\scshape\relax
  \ifnum\pdf@strcmp{\f@family}{\familydefault}=\z@
    \fontfamily{lmr}%
  \fi
  \fontshape\scdefault\selectfont}
\makeatother

\NewCommandCopy{\oldsqrt}{\sqrt} 
\renewcommand{\sqrt}[1][\ ]{%
  \def\DHLindex{#1}\mathpalette\DHLhksqrt}
\def\DHLhksqrt#1#2{%
  \setbox0=\hbox{$#1\oldsqrt[\DHLindex]{#2\,}$}\dimen0=\ht0
  \advance\dimen0-0.2\ht0
  \setbox2=\hbox{\vrule height\ht0 depth -\dimen0}%
  {\box0\lower0.71pt\box2}}

\let\oldBox\Box
\renewcommand\Box{\hspace{2cm}\oldBox}

\definecolor{kwmag}{rgb}{1., 0., 0.5}
\usepackage{sourcecodepro}
\usepackage{etoolbox}

\usepackage[binary-units,range-units=single]{siunitx}
\DeclareSIUnit[per-mode=reciprocal-positive-first]{\parsec}{pc}
\DeclareSIUnit[per-mode=reciprocal-positive-first]{\pixel}{pix}
\DeclareSIUnit[per-mode=reciprocal-positive-first]{\year}{yr}
\DeclareSIUnit[per-mode=reciprocal-positive-first]{\angstrom}{\text{Å}}
\DeclareSIUnit[per-mode=reciprocal-positive-first]{\Msun}{\ensuremath{M_\odot}}
\DeclareSIUnit[per-mode=reciprocal-positive-first]{\Lsun}{\ensuremath{L_\odot}}

\usepackage{cleveref}
\crefname{section}{\S}{\S\S}
\Crefname{section}{\S}{\S\S}
\crefname{subsection}{\S}{\S\S}
\Crefname{subsection}{\S}{\S\S}
\crefname{subsubsection}{\S}{\S\S}
\Crefname{subsubsection}{\S}{\S\S}
\crefname{figure}{Fig.}{Figures}
\Crefname{figure}{Fig.}{Figures}
\crefname{table}{Table}{Tables}
\Crefname{table}{Table}{Tables}
\crefname{equation}{Eq.}{Equations}
\Crefname{equation}{Eq.}{Equations}
\crefname{appendix}{Appendix}{Appendices}
\Crefname{appendix}{Appendix}{Appendices}
\crefname{chapter}{Chapter}{Chapters}
\Crefname{chapter}{Chapter}{Chapters}

\newcommand{\ReArc}{3.49}
\newcommand{\ReKpc}{2.15}
\newcommand{\bestBH}{9.21}
\newcommand{\bestBHLinear}{(1.62\spm{0.056}{0.054}) \times 10^9}
\newcommand{\bestQ}{0.3973}
\newcommand{\bestP}{0.9210}
\newcommand{\bestU}{0.9999}
\newcommand{\bestTH}{65.00}
\newcommand{\bestPH}{89.26}

\newcommand{\bestDM}{7.6211}
\newcommand{\bestDF}{1.2500}
\newcommand{\bestML}{2.335}
\newcommand{\onsBH}{0.0148}
\newcommand{\onsQ}{0.0135}
\newcommand{\onsP}{0.0156}
\newcommand{\onsU}{0.0042}
\newcommand{\onsDM}{-}
\newcommand{\onsDF}{0.2850}
\newcommand{\onsML}{0.0753}

\renewcommand*\copyright{{%
  \fontfamily{lmr}\selectfont\textcopyright}}

\title[SNELLS\textendash HD I]{SNELLS\textendash HD I: a first look at the stellar properties of the massive strong-lens galaxy \snl1\ with \SI{50}{\parsec} resolution}

\author[Poci \& Smith]{
Adriano Poci$^{1,2}$\thanks{E-mail: \href{mailto:adriano.poci@physics.ox.ac.uk}{adriano.poci@physics.ox.ac.uk}},
Russell J. Smith$^{2}$
\\
$^{1}$Astrophysics Sub-department, Department of Physics, University of Oxford, Keble Road, Oxford OX1 3RH, United Kingdom\\
$^{2}$Centre for Extragalactic Astronomy, University of Durham, Stockton Road, Durham DH1 3LE, United Kingdom}

\date{Accepted XXX. Received YYY; in original form ZZZ}

\pubyear{2024}

\begin{document}
\label{firstpage}
\pagerange{\pageref{firstpage}--\pageref{lastpage}}
\maketitle

\begin{abstract}
    We present a dynamical and chemical study of the centre of a massive early-type strong-lens galaxy ESO286-G022 (\snl1). Analysing new data obtained through the adaptive-optics-assisted Narrow-Field Mode of VLT/MUSE, we aim to measure the mass distribution and internal properties of \snl1\ at \(\sim\SI{50}{\parsec}\) resolution. In particular, we aim to address the tension in the reported IMF measurements of \snl1\ between strong-lens/dynamical and spectral-fitting techniques. We fit a triaxial orbital dynamical model to the measured stellar kinematics, including constraining the mass of the (resolved) central supermassive black-hole. The dynamical model is consistent with the mass-to-light ratio expected for a Kroupa-like IMF. We also employ a highly-flexible spectral-fitting technique, which instead favours a Salpeter-like IMF (low-mass slope \(\alpha\approx 2.3\)) over the same spatial region. To conclude, we discuss possible origins of this discrepancy, both intrinsic and technical.
\end{abstract}




\section{Introduction}\label{sec:intro}
Massive early-type galaxies (ETG) are the end product of galaxy evolution processes. As such, their physical properties encode information about said processes, and are therefore a useful tool for studying how galaxies form and evolve. In particular, the centre of ETG gives us access to a physical regime not found elsewhere. It is for this reason that they have been the focus of study for many years.\par
In general, the centres of massive ETG are characterised by extreme stellar densities, stellar metallicities, and old stellar populations, a dearth of gas, and dynamically-hot orbital configurations \citep[e.g.][]{dezeeuw1991,cappellari2016}. This is believed to be the result of an initial, intense burst of {\em in-situ} star-formation at early times, followed by more passive evolution through the accretion of lower-mass galaxies \citep[e.g.][]{naab2009,oser2010,zibetti2020}.\par
Observations of the centres of massive ETG provided indications that they formed with extreme stellar Initial Mass Functions (IMF), having an abundance of low-mass stars many times greater than what was found for the Solar neighbourhood --- a so-called `bottom-heavy' IMF \cite[e.g.][]{spiniello2011,dutton2012a,smith2015,lyubenova2016,labarbera2017,vandokkum2017,vaughan2018}. Yet concurrently, evidence was emerging that disfavoured such dwarf-rich IMF \citep[e.g.][]{brewer2012,davis2017,alton2018,collier2018}, using different techniques and galaxy samples between all of these works. This tension has been explicitly investigated with the analysis of \cite{smith2014}, which compared the different IMF measurement techniques on the same sample of galaxies, concluding that indeed the methodologies themselves produced incompatible results. A similar conclusion was drawn from the work of \cite{smith2015}. In order, therefore, to make progress on the intrinsic IMF variations within galaxies, these technical differences must first be understood.\par
One particular galaxy for which different measurement techniques provide inconsistent results is ESO286-G022 (\snl1\ hereafter), discovered as part of the SINFONI Nearby Elliptical Lens Locator Surveys \citep[SNELLS;][]{smith2015a}. Strong gravitational lensing, in combination with dynamical models, provide a mass-based estimate of the stellar IMF by marginalising over the dark matter (DM) contribution. For \snl1, this approach yields a stellar mass-to-light ratio (\mlStar) consistent with a \cite{kroupa2001} IMF \citep{newman2017}. Refining the lens model, \cite{collier2018a} found an `IMF mismatch' \(\alpha_{\rm IMF} = \Upsilon_{\rm lens} / \Upsilon_{\rm Kroupa} = 1.17\pm 0.09\) for lensing and reference mass-to-light ratios \(\Upsilon_{\rm lens}\) and \(\Upsilon_{\rm Kroupa}\), respectively. The \mlStar\ from the lensing is thus consistent with that predicted from a \cite{kroupa2001} IMF (with a slightly higher abundance of dwarf stars). While there is a range of possible values of \(\alpha_{\rm IMF}\) consistent with the lensing, dependent on the specific assumptions when separating the dark and baryonic mass, overall the lensing data are consistent with a relatively dwarf-poor IMF for \snl1\ \citep{newman2017}.\par
Conversely, flexible spectral-fitting techniques applied to optical absorption spectra of \snl1\ require IMF which are over-abundant in dwarf stars \citep{newman2017}. Not only is there a tension between these methodologies, then, but it is especially difficult to reconcile the fact that the spectroscopic results favour heavier populations than inferred by the dynamics (even with its DM uncertainty). \snl1\ therefore represents an interesting test case on which to refine these modelling techniques in order to ascertain the cause of this IMF discrepancy.\par
We have previously modelled spatially-resolved integral-field unit (IFU) data of \snl1\ from Very Large Telescope (VLT)/ Multi-Unit Spectroscopic Explorer (MUSE), taken in the wide-field mode (WFM). From the dynamical model in that work \citep{poci2022}, \snl1\ appears to be considerably rotationally-supported despite its global morphology and high velocity dispersion. There was evidence of mild triaxiality, and mostly isotropic stellar orbits. That work also reaffirmed the high degree of compactness of \snl1. One pertinent result from that model was that the dynamics and lensing, when computed independently of one another, show excellent agreement in the enclosed mass. We can hence consider the constraints from lensing and dynamics as consistent, to be contrasted with the discrepant spectroscopic results.\par
In this work, we explore the central region of \snl1\ using new high-resolution data --- some of the highest physical resolution stellar kinematics published to date for a galaxy outside the Local Group. We aim to characterise the dynamics and populations using an array of sophisticated modelling techniques in order to understand how the different mass components are distributed. Ultimately, we aim to discern the cause of the tension between the different modelling techniques, and measure the intrinsic IMF of \snl1. Throughout this work, we assume a Planck 2018 cosmology \citep{planckcollaboration2020a}, with \(H_0 = \SI{67.66}{\kilo\metre\per\second\per\mega\parsec}\) and \(\Omega_m = 0.3111\).

\section{Data \& Target}\label{sec:data}
\begin{table}\label{tab:props}
    \begin{threeparttable}
        \caption{Physical properties of \snl1.}
    \newcolumntype{L}{>{$}l<{$}} 
    \begin{tabular}{r@{\hspace{0.9\tabcolsep}}L|L|c}\hline\hline
        Redshift & z & 0.0312 & \text{\cite{smith2015a}}\\\hline
        \multirow{2}{*}{Einstein radius} & \multirow{2}{*}{\(\Rein\)} & 2.38\si{\arcsecond} & \multirow{2}{*}{\cite{smith2015a}}\\
         && 1.48\ \si{\kilo\parsec} &\\\hline
        Total mass & \logM & 10.98 & \text{\cite{collier2018a}}\\\hline
        \multirow{2}{*}{Distance} & \multirow{2}{*}{\(D^\dagger\)} & 132\ \si{\mega\parsec} & luminosity\\
         && 128\ \si{\mega\parsec} & angular diameter\\\hline
        \multirow{2}{*}{Effective radius} & \multirow{2}{*}{\(R_\eff^{F814W}\)} & \ReArc\si{\arcsecond} & \multirow{2}{*}{\cite{poci2022}}\\
         && \ReKpc\ \si{\kilo\parsec} & \\\hline
    \end{tabular}
  \begin{tablenotes}
    \footnotesize
    \item[$^{\dagger}$] Derived from the redshift for our assumed cosmology.
  \end{tablenotes}
    \end{threeparttable}
    \end{table}
In this work we present new observations of a well-studied, relatively nearby strong-lens galaxy \snl1. The physical properties of \snl1\ are summarised in \cref{tab:props}. \snl1\ is acting as a strong-lens to a background source galaxy at \(z=0.926\); the lens modelling and results are described in detail in \cite{smith2015a} and \cite{collier2018a}. This galaxy has been studied in the context of the stellar IMF thanks to the robust enclosed-mass constraints provided by the lensing \citep{newman2017}. Despite being a massive ETG, the lens model results in a \(\mlStar=4.61 \pm 0.39\) which is consistent with a Milky-Way (MW)-like IMF \citep{kroupa2002}. In stark contrast, spectral analyses of this galaxy imply heavily dwarf-rich stellar populations, consistent with galaxies of similar morphology and central velocity dispersion.
\subsection{Spectroscopy}\label{ssec:spec}

\begin{figure}
    \centerline{
        \includegraphics[width=\columnwidth]{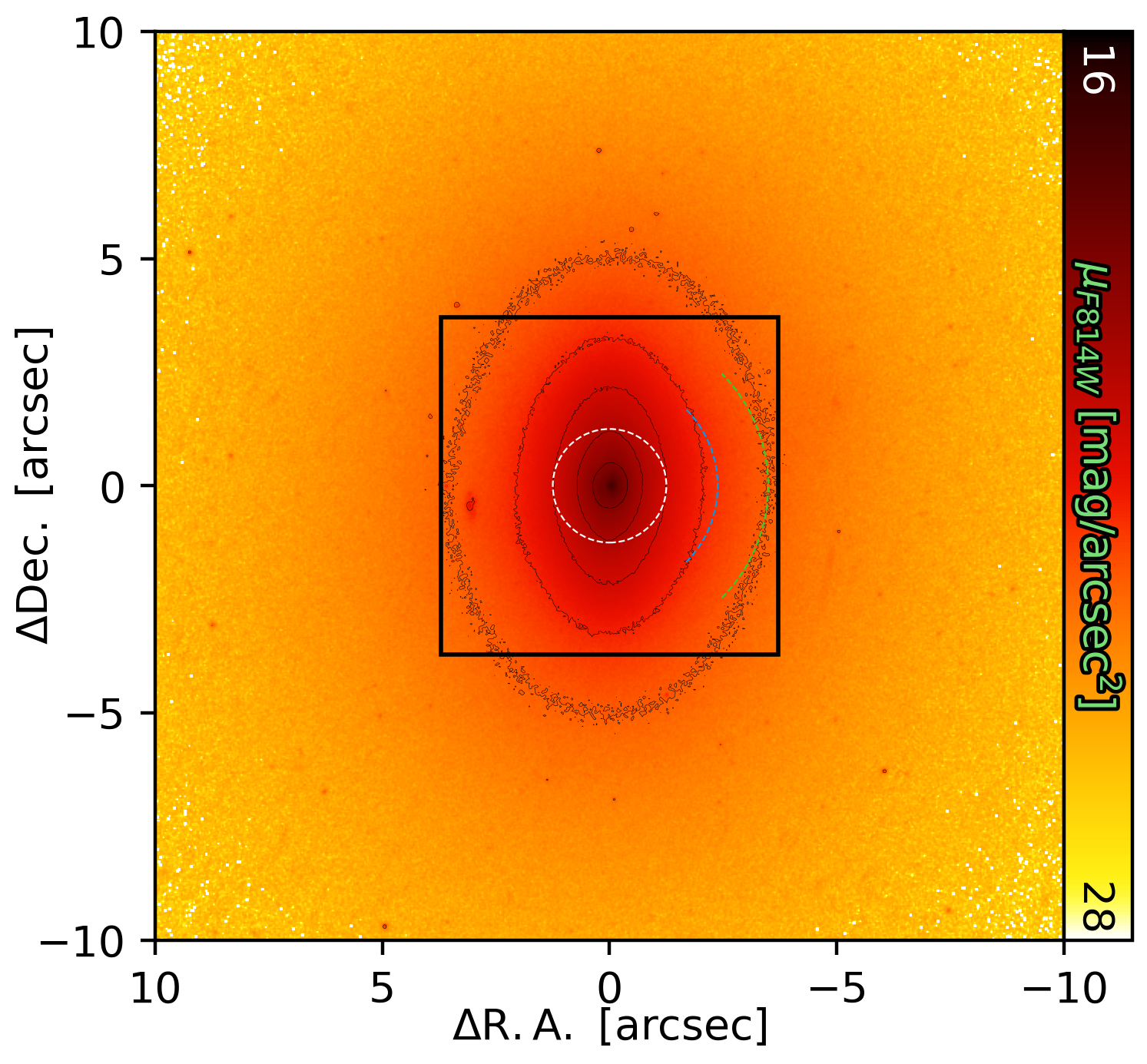}}
    \caption{{\em HST} \(F814W\) image of \snl1. Isophotes are shown as thin black lines. The solid black box shows the FoV of the MUSE NFM. The dashed white circle has a diameter of \(\qty{1.25}{\arcsecond}\), which is the region explored in this work. The blue and green dashed arcs demarcate \(\Rein\) and \(R_\eff\), respectively. \snl1\ exhibits flattened isophotes, though the effect of the dust lane is visible in the inner-most contour.}
    \label{img:photo}
\end{figure}

With this work, we present new MUSE \citep{bacon2010} observations in the Narrow-Field Mode (NFM) with adaptive optics \citep[AO;][]{arsenault2008,strobele2012} under program ID 109.22X3.001. The NFM of MUSE has a field-of-view (FoV) of \(7.5\si{\arcsecond}\), sampled with \(0.025\si{\arcsecond}\) pixels. The spectral range is \(\qtyrange[range-phrase=-]{4800}{9300}{\angstrom}\), with a notch filter across \(\qtyrange[range-phrase=-]{5780}{6050}{\angstrom}\) to block reflected light from the lasers of the Laser Tomography AO system. \snl1\ was observed in \(6\) observing blocks, each with a different dithering pattern and separate off-source sky frames, with a total on-source integration of \(\SI{14973}{\second}\). The raw frames were reduced by the standard ESO pipeline. \cref{img:photo} shows {\em Hubble Space Telescope} ({\em HST}) \(F814W\)-band photometry of \snl1\ from \cite{smith2015a}, overlaid with the footprint of the MUSE NFM.\par
The small pixel scale of NFM data results in significant read-out noise, meaning that the signal-to-noise ratio \((S/N)\) of the source drops off rapidly from the centre. Combined with some sky-subtraction issues in the data (see \cref{app:sky}), we choose to analyse the NFM data only in the brightest region. The high signal in this region means that those pixels are robust against sky-subtraction peculiarities, as well as requiring less spatial binning to reach the high signal-to-noise \((S/N)\) used in our analysis. It is also the critical region of interest, and one which is uniquely resolved by the NFM. Thus, we exclude all data outside of a circular aperture with diameter \(\sim\)\SI{1.25}{\arcsecond} \((\sim \SI{805}{\parsec})\) centred on the brightest pixel. Regions outside this aperture are resolved by other data-sets, such as the MUSE WFM in \cite{poci2022}. A joint analysis will be explored in a forthcoming work.\par
The point-spread function (PSF) of the NFM case is complicated by the use of the AO system. Such systems are generally expected to be characterised by a superposition of two Moffat profiles; one describing the corrected core, and one to account for the uncorrected natural seeing \citep[e.g.][]{fetick2019}. Since our target covers the full FoV of the NFM, we can not estimate the PSF from the science observations as there are no isolated stars in the field. Thus, we have instead estimated the PSF by fitting a model to the predicted PSF from the MUSE exposure-time calculator (ETC)\footnote{\url{https://www.eso.org/observing/etc/bin/gen/form?INS.NAME=MUSE+INS.MODE=swspectr}}.\par
The PSF was requested under the observational setup used in Phase 2; namely, with a maximum airmass of \(1.5\)\footnote{The model for the NFM PSF from the MUSE ETC does not currently depend on the turbulence, and thus neither does our modelled PSF.}. The NFM AO performance depends sensitively on airmass, ambient seeing and atmospheric coherence time (MUSE manual v12.1). Our six observations were acquired over three nights at low airmass (\(1.04 - 1.26\), median \(1.07\)), with good seeing (\(\qtyrange[range-phrase=-]{0.42}{0.76}{\arcsecond}\), median \(\SI{0.49}{\arcsecond}\)) and coherence time (\(\qtyrange[range-phrase=-]{3.5}{7.4}{\milli\second}\), median \(\SI{5.5}{\milli\second}\)). According to the MUSE manual v12.1, Figures 37-39, these conditions are usually sufficient to yield close to the optimal delivered PSFs, and hence we expect that adopting the ETC model should not be an unreasonably optimistic assumption.\par
\begin{figure}
    \centerline{\includegraphics[width=\columnwidth]{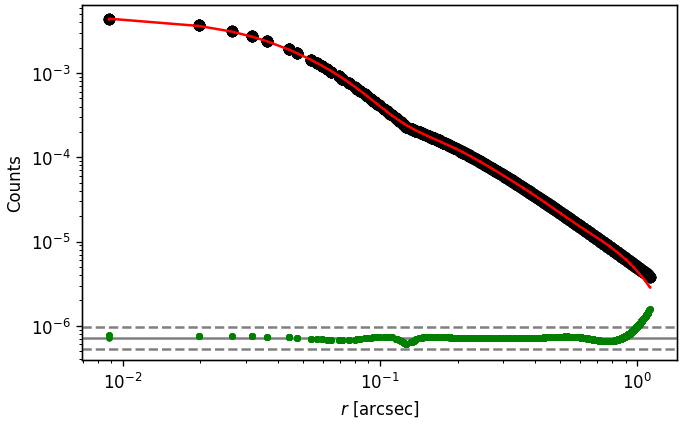}}
    \caption{One-dimensional azimuthally-averaged brightness profile of the model PSF. Black points show the PSF `image' (the model provided by the MUSE ETC). The red solid line shows the MGE model fit to that image. The green points show the residuals of the fit \(({\rm image-model/image})\), arbitrarily offset for presentation. The grey solid and dashed lines show the \(0\) and \(\pm10\%\) (on a linear axis) of the residuals, respectively.}
    \label{img:PSF}
\end{figure}

We chose to evaluate the PSF at a reference wavelength of \(\lambda_{\rm ref} = \SI{6500}{\angstrom}\), as a compromise over the full spectral range from which we measure the stellar kinematics (\cref{sec:stelkin}). We fit a multi-Gaussian Expansion (MGE) model \citep{cappellari2002} to the PSF image. This is because not only is MGE general enough to accurately reproduce a double Moffat profile expected for the NFM PSF, but also because this parametrisation is used later by our dynamical modelling implementation (\cref{ssec:schw}). The PSF fit is shown in \cref{img:PSF}, and the model is tabulated in \cref{tab:mgePSF}.\par
At the angular diameter distance of \snl1, \(\SI{1}{\arcsecond} \approx \SI{0.6}{\kilo\parsec}\). Accounting for the PSF, where the vast majority of the weight of the PSF core is \(\lesssim \SI{0.06}{\arcsecond}\), these observations have an effective resolution of \(\sim \SI{36}{\parsec}\). Hence, even the restricted aperture we've imposed is resolved by \(\sim 20\) resolution elements (PSF cores) across.
\begin{table}
    \begin{tabular}{S[table-format=1.4]|S[table-format=1.4]|S[table-format=1.4]}
        {$\bar{f}$} & {$\sigma\ [\si{arcsec}]$} & {${\rm FWHM}\ [\si{arcsec}]$}\\
       \hline
       0.4239 & 0.0219 & 0.0516\\
       0.5045 & 0.0434 & 0.1022\\
       0.0445 & 0.1104 & 0.2600\\
       0.0216 & 0.2107 & 0.4962\\
       0.0055 & 0.5392 & 1.2698\\
       \hline
       \end{tabular}
   \caption{PSF MGE model for the NFM observations, showing the normalised amplitude ({\em left}) and width ({\em right}) of each Gaussian. The model is normalised such that \(\sum f_j = 1\) for \(j\) MGE Gaussians. We fit a circular PSF model, and so the Gaussian components all have an axis ratio of \(1\).}
   \label{tab:mgePSF}
\end{table}
\subsubsection{Non-Stellar Features}
\begin{figure}
    \centerline{
        \includegraphics[width=\columnwidth]{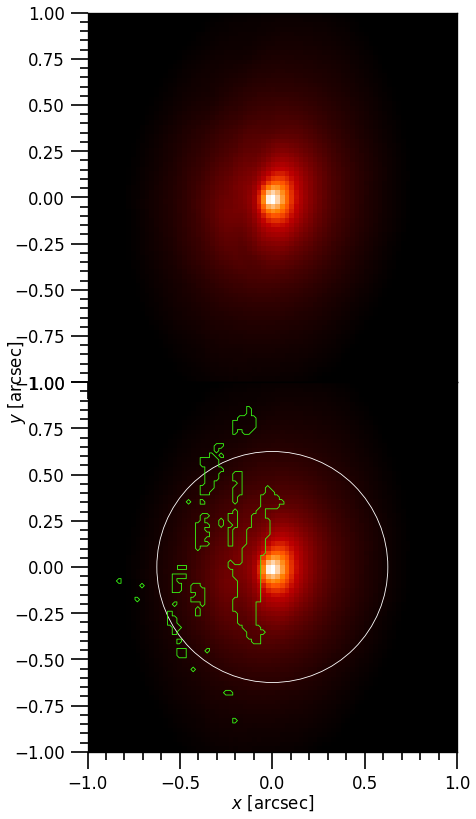}}
    \caption{{\em Top:} Synthetic {\em HST} \(F814W\)-band image of \snl1. Multiple dusty structures are visible on the near side \snl1. {\em Bottom:} Identical image as above, but with the derived mask overlaid in green showing where dust has been detected, and subsequently removed. Overlaid in white is the circular FoV over which the spectroscopy is analysed in this work.}
    \label{img:dust}
\end{figure}

The NFM data show attenuation of the stellar light by dust. Nuclear bifurcated dust lanes are clearly seen in the white-light (spectrally-integrated) data-cube. The smallest resolved dust structure, in a clear disk/ring morphology, has an approximate radius of \(\SI{0.4}{\arcsecond}\ (\SI{260}{\parsec})\). Another dust structure is seen only on the approaching side of the galaxy (at least down to the depth of these data), extending out to \(\sim\SI{2}{\arcsecond}\ (\SI{1.3}{\kilo\parsec})\) in an arc. \cref{img:dust} shows a synthetic \(F814W-\)band image from the NFM data-cube, scaled to highlight the central dusty ring structure.
We mask the dust lanes henceforth since both the stellar population and kinematic measurements could be biased in those sight-lines. The masking of dusty pixels was done in the following way. First, a synthetic {\em HST} Wield-Field Camera 2 \(F439W-F814W\) colour image was constructed to maximise contrast of the dust lane within the MUSE wavelength range. This was achieved by applying the respective filter curves directly to the spectra of the NFM data-cube, then taking the difference. The synthetic colour image was then unsharp-masked using a Gaussian filter with a kernel of \(1.5\) pixels. From that unsharp-masked image, all pixels below a threshold of \(290\) were assigned to the mask. This value was chosen to conservatively include all the foreground dust-obscured regions.

\section{Central Stellar Kinematics at Parsec-Scale Resolution}\label{sec:stelkin}
The central regions of ETG have been of interest in galaxy-evolution studies for decades, believed to host a plethora of interesting and still-debated astrophysical processes. The central region of \snl1\ appears to be especially interesting even in the context of similarly-massive ETG. The rotation in \snl1\ (visible already from the long-slit data of \citealt{newman2017} and the MUSE WFM; \citealt{poci2022}) is unusually high given its morphology and mass. In this section we look at measuring stellar kinematics at the high spatial resolution afforded by the NFM to study how the centre of this galaxy is gravitationally supported.\par
Before fitting the spectra to measure the kinematics, we spatially-bin the data-cube using a Python implementation\footnote{Available at \url{https://pypi.org/project/vorbin/}} of the Voronoi binning algorithm \citep{cappellari2003} to a target \(S/N=40\). This \(S/N\) allows us to take advantage of the high spatial resolution of the NFM, while still being sufficient to measure non-Gaussian shapes of the line-of-sight velocity distribution (LOSVD). We consider the full rest-frame spectral range of MUSE covering \(\lambda \in [\numrange[range-phrase=\ensuremath{,{}}]{4750}{9000}]\ \si{\angstrom}\). We mask inadequately-treated telluric features over the range \([\numrange[range-phrase=\ensuremath{,{}}]{7600}{7700}]\ \si{\angstrom}\), with an additional mask over the range \([\numrange[range-phrase=\ensuremath{,{}}]{5700}{6100}]\ \si{\angstrom}\) to account for the notch filter of the AO system.\par
Stellar kinematics were measured using the {\sc pPXF} Python package\footnote{Available at \url{https://pypi.org/project/ppxf/}} \citep{cappellari2004, cappellari2017}, and with the X-Shooter Stellar Library \citep[XSL;][]{verro2022a} in order to cover the full MUSE wavelength range. The library spectra were convolved with the wavelength-dependent line-spread function of MUSE prior to fitting. To robustly characterise the absorption-line shapes, and thus the LOSVD, a \(3\)rd-order additive polynomial was used during fitting, while a \(3\)rd-order multiplicative polynomial was included to account for low-frequency continuum mismatch. The LOSVD in each spatial bin is parametrised by four Gauss-Hermite coefficients \citep{vandermarel1993}, approximately its mean velocity, velocity dispersion, skewness, and kurtosis, respectively. The measured kinematics can be seen in the left column of \cref{img:schw}.\par
\begin{figure}
    \centerline{\includegraphics[width=\columnwidth]{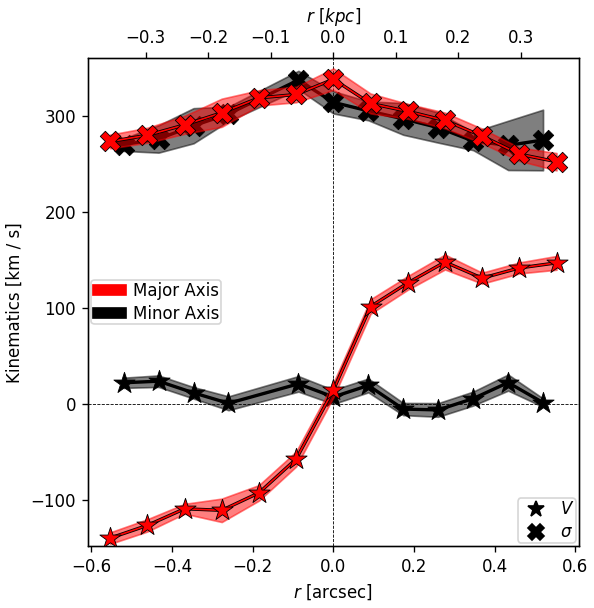}}
    \caption{Pseudo-slit profiles of the stellar velocity (star symbols) and velocity dispersion (cross symbols) for \snl1. The profiles are extracted within pseudo-slits laid on the binned kinematic FoV \((S/N=40)\) with a width of \(\pm\qty{0.2}{\arcsecond}\). Profiles are shown for major- ({\em red}) and minor- ({\em black}) axis slits. Shaded regions illustrate the mean measurement uncertainty of the kinematic fits in each region.}
    \label{img:slits}
\end{figure}

The NFM covers the high-rotation region which was known from the WFM, which exhibits a maximum amplitude of \(\pm \SI{245}{\kilo\metre\per\second}\) within \(\SI{3}{\kilo\parsec}\). However, in the NFM, the rotation extends further towards the centre than was previously seen, and becomes more confined to the major axis plane, in contrast to the dynamically-hotter large-scale rotation. Combined with the dust geometry, this is strong evidence of a nuclear stellar disk, which is co-spatial with the dust and ionised gas disks (\cref{sec:gas}). In the stellar component, this disk exhibits a maximum rotation amplitude of \(\pm \SI{165}{\kilo\metre\per\second}\) within \(\SI{260}{\parsec}\). We present pseudo-slit profiles of the stellar velocity and velocity dispersion in \cref{img:slits}. This figure also re-iterates that \snl1\ is pressure-supported; the \(|V|/\sigma\) can be read directly from \cref{img:slits} as \(\sim 0.5\).
\subsection{Orbital Dynamical Models}\label{ssec:schw}
We employ highly-general triaxial \shw\ orbit-superposition models \citep{schwarzschild1979,vandenbosch2008} in order to understand the internal structures which give rise to the new observed kinematics. We have already fit \shw\ models for \snl1\ \citep{poci2022} to the WFM, though were unable to confidently disentangle the mass components in the centre due to the relatively large PSF. With a dynamical model of the new kinematics, we aim to break the degeneracy of the central mass composition, since the central supermassive black-hole (SMBH) is dynamically resolved by these data.\par
Briefly, the \shw\ models integrate, for a given gravitational potential, a large library of stellar test orbits which are permitted to reside in said potential. From this library, the optimal subset is fit for in order to most accurately reproduce the observed stellar kinematics. This entire process is repeated for many choices of intrinsic (de-projected) gravitational potentials which are all consistent with the observed (projected) stellar mass distribution. We take the model with the best goodness-of-fit measure, and corresponding best-fit subset of orbits, to represent the intrinsic mass distribution and stellar kinematics of the galaxy.\par
The model we use for the projected mass distribution (in order to sample plausible intrinsic gravitational potentials) is derived by fitting a multi-Gaussian Expansion \citep[MGE;][]{cappellari2002} to the \(HST\ F814W\) photometry. By using the stellar light to model the projected mass, we implicitly assume that the \mlStar\ is spatially-constant (over the small FoV of the kinematics). We explore the impact of this assumption in a forthcoming work. In the general case, the intrinsic gravitational potentials are characterised by \(7\) free parameters: \begin{itemize*}[label={---}] \item the mass of the central SMBH, \(M_\bullet\) \item the axis ratios of the intrinsic mass distribution, \(q, p, u\), where \(q = C/A\), \(p = B/A\), and \(u = A^\prime/A\), for major, intermediate, and minor axes \(A\), \(B\), and \(C\), and projected major axis \(A^\prime\) \item parameters of the Navarro-Frenk-White \citep[NFW;][]{navarro1996} cold DM halo, the concentration \(C_{\rm DM}\), and mass fraction at \(r_{200}\), \(M_{200}/M_\star\) \item a spatially-constant \mlStar\ scale factor, \(\Upsilon\) \end{itemize*}. In this work, due to the restricted FoV of the stellar kinematics, we do not attempt to fully constrain the DM halo, but instead tie the two parameters together using the concentration-mass relation of \cite{dutton2014a}, so that in practise, we sample only \(M_{200}/M_\star\) and are therefore reduced to \(6\) free parameters.\par
The implementation of the \shw\ model used in this work samples three conserved integrals-of-motion which are, in a triaxial potential, the energy \(E\), a non-analytic \(z\)-axis angular momentum \(I_2\), and a third non-classical conserved integral \(I_3\). We sampled \(E\), \(I_2\), and \(I_3\) in \(30\) logarithmic, and \(25\) and \(18\) linear steps, respectively, generating a library of tube orbits. The same set-up was used to generate a separate library of box orbits. We applied a dithering to this orbital sampling of a factor of \(3\) (in each integral; \(3^3\) in total). For every starting coordinate in the integral-of-motion space, this dithering samples \(3\) slightly perturbed coordinates in each dimension, producing a `cloud' of orbits around each point, and reducing discreteness in the orbit library.\par
\begin{figure}
    \centerline{
        \includegraphics[width=\columnwidth]{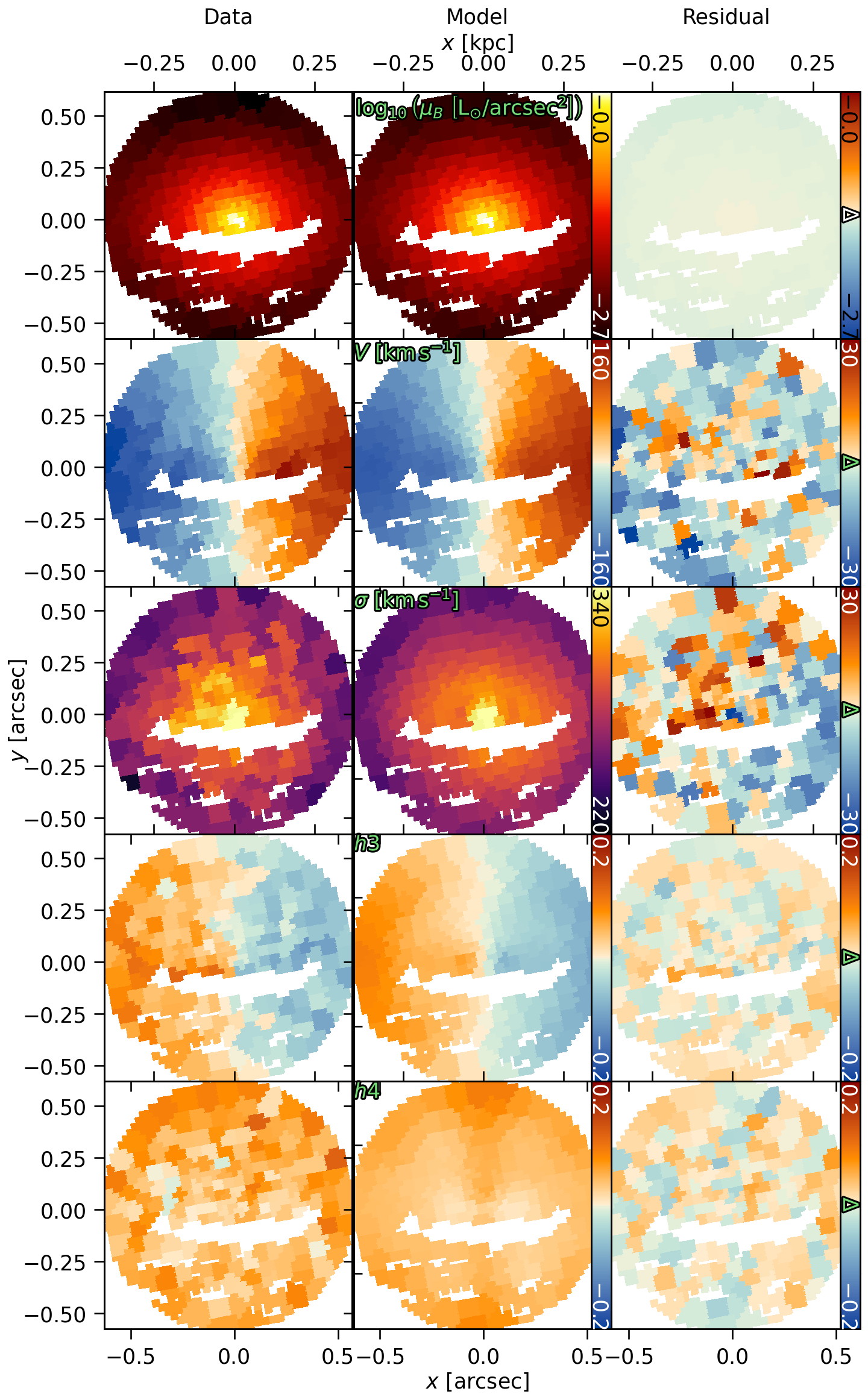}}
    \caption{The measured stellar kinematics ({\em left}), the corresponding best-fit \shw\ model ({\em centre}), and the residuals (data\(-\)model; {\em right}) for the NFM data of \snl1. The rows from top to bottom show the projected surface brightness, rotational velocity, velocity dispersion, \(h_3\), and \(h_4\).}
    \label{img:schw}
\end{figure}
\begin{figure*}
    \centerline{\includegraphics[width=\textwidth]{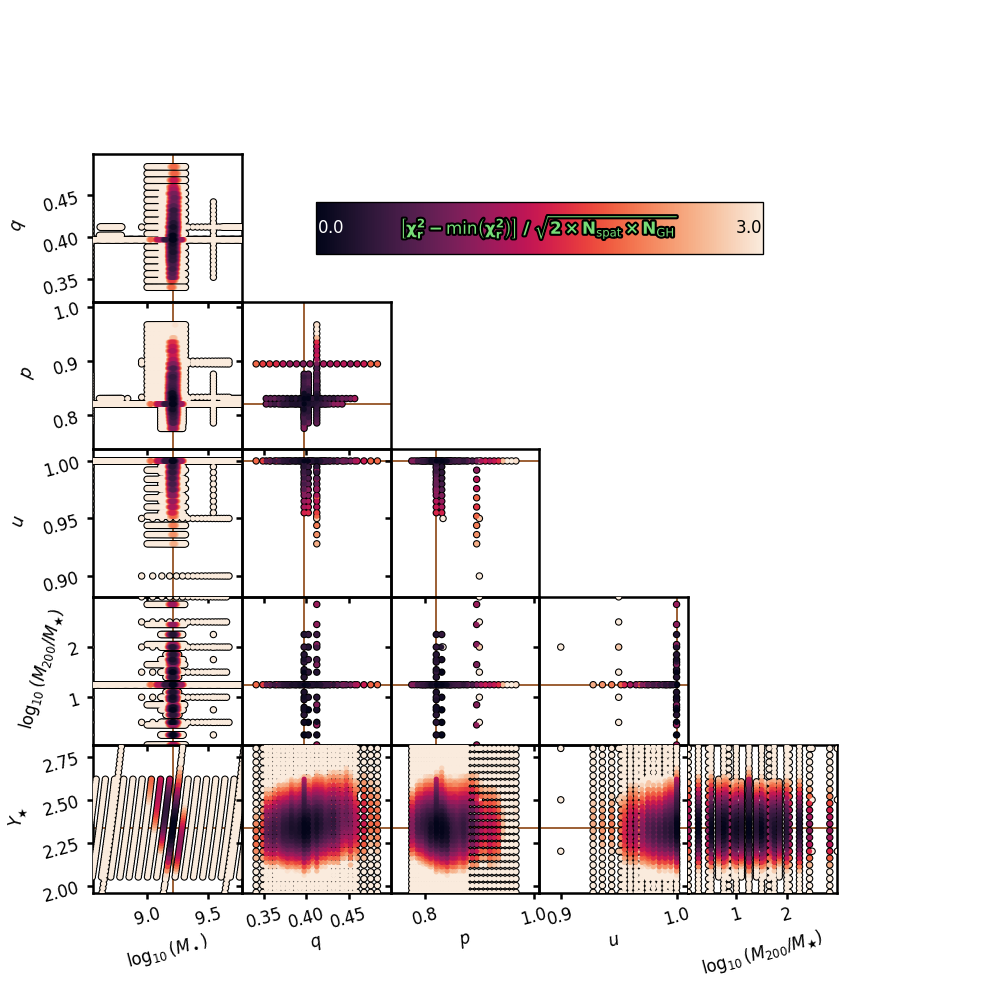}}
    \caption{Subset of the parameter exploration for the \shw\ model of \snl1. Points are coloured by relative goodness-of-fit as given by \cref{eq:chi2}. Best-fit parameters are demarcated by solid brown lines. The panels show a subset of the total parameter-space explored, focusing on the low \(\chi^2\) region. The full parameter-space exploration is shown in \cref{app:shw}.}
    \label{img:shwcorn}
\end{figure*}
\begin{figure}
    \centerline{\includegraphics[width=\columnwidth]{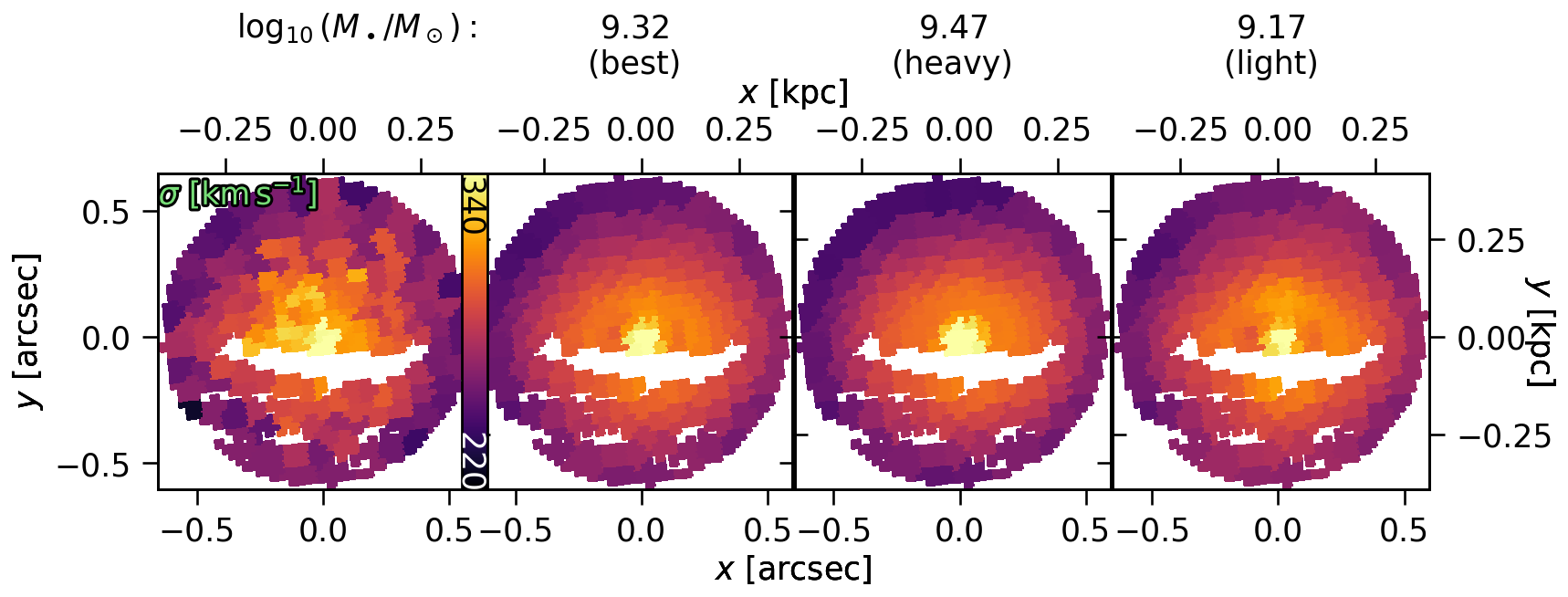}}
    \caption{Impact of changing the SMBH mass on the predicted stellar velocity dispersion. From left to right are the measured kinematics, the best-fit model, a model with an over-massive \(M_\bullet\), and a model with an under-massive \(M_\bullet\), given by adjacent steps in the parameter-space. The other parameters of the model are held fixed to the best-fit values. The impact of changing \(M_\bullet\) is seen clearly in the velocity dispersion, where the central peak is visibly over- and under-predicted for the over- and under-massive models, respectively.}
    \label{fig:bhModels}
\end{figure}

At every iteration, the model `observables' were first spatially binned and convolved with the PSF of the NFM observations in order to be directly comparable to our measured stellar kinematics. We then defined the best-fitting model as that which has the minimum \(\chi^2\) across all four Gauss-Hermite kinematic moments. The best-fitting \shw\ model is shown in the middle column of \cref{img:schw}, with the residuals (data\(-\)model) in the right column. The exploration of the model parameters is shown in \cref{img:shwcorn}.
Models are coloured by their goodness-of-fit, relative to the best-fitting model, as
\begin{flalign}
    && \left[\chi_r^2 - \mathrm{min}(\chi_r^2)\right] \big/ \sqrt{2 N_{\rm obs} N_{\rm GH}} &&\label{eq:chi2}
\end{flalign}
for the number of Voronoi bins, \(N_{\rm obs}=179\), the number of Gauss-Hermite kinematic moments fit in the \shw\ model, \(N_{\rm GH}=4\), and \(\chi_r^2\) as defined in \cite{zhu2018a,poci2019}.
\cref{img:shwcorn} indicates that the constraints on the DM are weak, as might be expected given that the data cover only the most baryon-dominated region. Conversely, we see that the constraints on the SMBH mass are much more secure. The best-fitting SMBH mass is \(M_\bullet = \bestBHLinear\ \si{\Msun}\). Despite the grid-based approach to our parameter sampling, we still resolve the \(M_\bullet-\Upsilon_\star\) anti-correlation. \cref{fig:bhModels} shows the impact of changing \(M_\bullet\) on the stellar kinematics.\par
\begin{figure}
    \centerline{\includegraphics[width=\columnwidth]{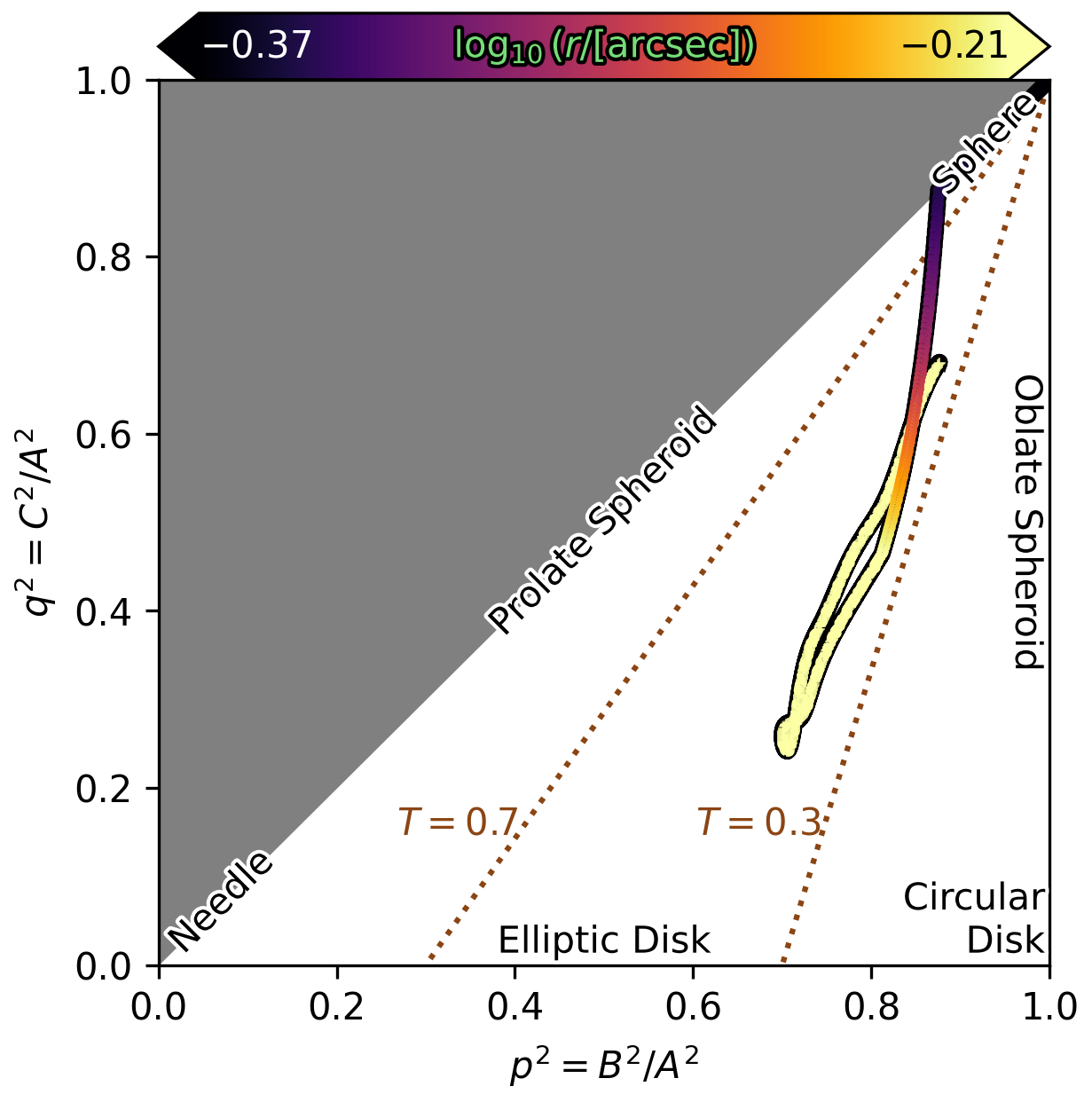}}
    \caption{Minor-to-major axis ratio \(q\) against intermediate-to-major axis ratio \(p\), as a function of radius. The left and right extrema of the colourbar denote the FWHM of the PSF and maximum extent of the kinematics, respectively. The model extends beyond these bounds, but is unconstrained by the spectroscopy in those regions.}
    \label{img:axis}
\end{figure}

From \cref{img:schw}, the model is able to reproduce all the measured moments of the LOSVD to a high degree of accuracy. The physical properties of the best-fitting model are given in \cref{tab:snl1}.
\begin{table}
    \centerline{
    \begin{tabular}{c|c|c|c}
        {\bf Parameter} & {\bf Description} & {\bf Best} & {\bf \(1\sigma\)}\\\hline\hline
        \(\log_{10}(M_\bullet/\si{\Msun})\) & Black-Hole Mass & \(\bestBH\) & \(\onsBH\)\\\hline
        \(q\) & Intrinsic Shape & \(\bestQ\) & \(\onsQ\)\\
        \(p\) & Intrinsic Shape & \(\bestP\) & \(\onsP\)\\
        \(u\) & Intrinsic Shape & \(\bestU\) & \(\onsU\)\\\hline
        \(\theta^\prime\) & Viewing Angle & \(\bestTH\si{\degree}\)&\\
        \(\phi^\prime\) & Viewing Angle & \(\bestPH\si{\degree}\)&\\
        \(\psi^\prime\) & Viewing Angle & \(\bestPH\si{\degree}\)&\\\hline
        \(C_{\rm DM}\) & DM Concentration & \(\bestDM\) & \(\onsDM\)\\
        \(\log_{10}\left(M_{200}/M_\star\right)\) & DM ratio at \(r_{200}\) & \(\bestDF\) & \(\onsDF\)\\\hline
        \(\Upsilon\; [\si{\Msun/\Lsun}]\) & Global \(M/L\) & \(\bestML\) & \(\onsML\)
    \end{tabular}
    }
    \caption{Free parameters of the \shw\ model, their best-fitting values, and the associated uncertainties. Uncertainties are derived by taking the spread of parameter values of all models within \(1\sigma\) of the best-fitting solution, which is visualised in \cref{img:shwcorn}. Note that \(\theta^\prime, \phi^\prime, \psi^\prime\) are derived from the best-fitting \(q,p,u\), while \(C_{\rm DM}\) is computed from \(M_{200}/M_\star\) (\cref{ssec:schw}).}
    \label{tab:snl1}
\end{table}
The dynamical model suggests an inclination of \(\SI{65}{\degree}\), which implies that the intrinsic thickness of the nuclear disk is thinner still than what appears in projection in the kinematic maps. Indeed, we find a remarkably flattened structure in the central region. This is perhaps expected, given the high degree of ordered rotation, but is atypical for galaxies of this mass and global morphology. The minor-to-major axis ratio \(C/A\) becomes as flat as \(\sim 0.5\) towards the edge of the FoV. \cref{img:axis} presents the axis ratios as a function of radius.
It clearly illustrates that the nuclear disk is truly a distinct structure, since at both smaller and larger radii, \snl1 is more spherical than in this region \((\sim \SI{0.4}{\arcsecond}/\SI{260}{\parsec})\). The origin of this disk, though discussed further below, remains unclear. However, on-going work with multiple tracers (including ionised and molecular gas) aims to solve this puzzle.\par
\begin{figure}
    \includegraphics[width=\columnwidth]{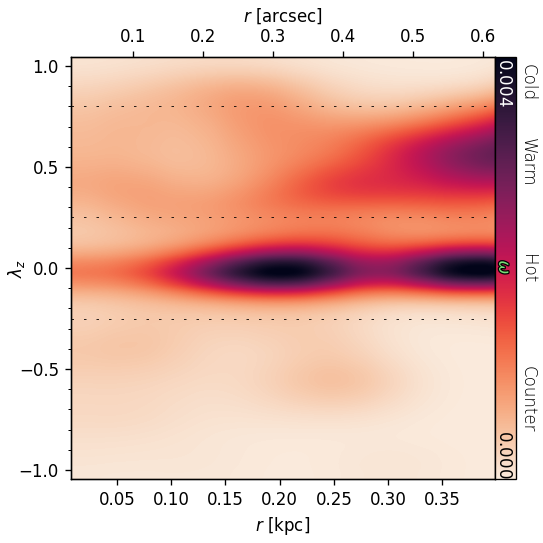}
    \caption{Orbital circularity \(\lambda_z\) as a function of time-averaged mean orbital radius, for the best-fit \shw\ model. Darker colours indicate higher contributions to the model. Horizontal dashed lines demarcate the orbital categories defined in \protect\cite{zhu2018a}, with corresponding labels on the right.}
    \label{img:circ}
\end{figure}
The \shw\ model provides a measure of the intrinsic dynamical support within \snl1. \cref{img:circ} shows the orbital circularity from the fit to the NFM kinematics, defined in \cite{zhu2018a}, and given here in \cref{eq:circ}.
\begin{flalign}\label{eq:circ}
    \lambda_z &= \overline{L_z} \Large/ \left( \overline{r} \cdot \overline{V_2} \right) &&&&\\
    \text{with}&& & &&\tag*{}\\
    \overline{L_z} &= \overline{x V_y - y V_x} &&&&\tag*{}\\
    \overline{r} &= \sqrt{\overline{x^2 + y^2 + z^2}} &&&&\tag*{}\\
    \overline{V_2}^2 &= \overline{V_x^2 + V_y^2 + V_z^2 + 2 V_x V_y + 2 V_x V_z + 2 V_y V_z} &&&&\tag*{}
\end{flalign} The orbital circularity is a normalised measure of the vertical angular momentum. According to this definition, \(\lambda_z = 1\ (-1)\) corresponds to co-rotating (counter-rotating) disk orbits, while \(\lambda_z=0\) corresponds to dynamically-hot/box/chaotic orbits. The peaks in the circularity distribution in \cref{img:circ} can be linked to the structures already identified in the kinematics. A heavily pressure-supported component is clearly seen at \(\lambda_z=0\), which dominates the mass budget. There exists a weak peak at high circularity \((\lambda_z\approx 1)\) and small radius \((r\approx \SI{0.15}{\kilo\parsec})\), which represents the nuclear disk. A warmer component \((\lambda_z\approx 0.5)\) begins to emerge towards the edge of the FoV, which was found in \cite{poci2022} to be responsible for the larger-scale rotation seen in the WFM.\par
The limited extent of the NFM data might result in the under-population of particular orbital families in the \shw\ model \citep[e.g.][]{krajnovic2008}. While in this work we do not attempt to combine the NFM and WFM kinematics, we can broadly compare the individual model outputs, though some are not directly comparable. We also re-emphasise here that there is no spatial overlap between the two data-sets, so direct quantitative comparisons would require an extrapolation in one direction. From the NFM model, we found \((q, p, u) = (\bestQ, \bestP, \bestU)\), compared to the WFM model with \((q, p, u) = (0.4949, 0.8390, 0.9910)\). This indicates a preference for rounder shapes in the WFM, but mild triaxiality in both. The models in fact independently predict that the central region, around the extent of the nuclear disk, is flatter than the outskirts. Given this region is unresolved in the WFM, it would naturally result in a thicker average intrinsic shape. The comparison of the parameters of the DM halo is not as straight-forward. For the NFM, we do not expect to constrain those parameters, and thus decided to couple \(C_{\rm DM}\) and \(\log_{10}(M_{200}/M_\star)\). In the WFM model, these parameters could vary independently. We find that the NFM model predicts a smaller fraction of dark matter even at fixed radius compared to the WFM model. The comparison of \(\Upsilon_\star\) is also not trivial, since the WFM model included spatially-varying \mlStar\ during the construction of the gravitational potential, which re-normalises the output \(\Upsilon_\star\).\par
We can instead compare the total masses predicted by the models at fixed radii. The NFM model produces an enclosed stellar mass of \(\log_{10}\left[M_\star(<\SI{1}{\kilo\parsec})\right] = 10.65\) (emphasising that this is an {\em extrapolation} beyond the NFM data), while the WFM model produces \(\log_{10}\left[M_\star(<\SI{1}{\kilo\parsec})\right] = 10.77\). Similarly, the total enclosed masses are \(\log_{10}\left[M_{\rm dyn}(<\SI{1}{\kilo\parsec})\right] = 10.67\) and \(\log_{10}\left[M_{\rm dyn}(<\SI{1}{\kilo\parsec})\right] = 10.81\) for the NFM and WFM models, respectively. Thus, in lieu of the multi-scale model, we conclude that the NFM model alone is not significantly biased in the physical properties we analyse in this work.\par

\section{Central Stellar Populations at Parsec-Scale Resolution}
Spectroscopic IMF measurements are often reported to become dramatically more dwarf-rich towards the centres (within the central \(\sim \SI{1}{\kilo\parsec}\)) of similar galaxies \citep[e.g.][]{labarbera2017,labarbera2019,martin-navarro2021a}. \snl1\ exhibits similarly dwarf-rich spectral signatures in its centre \citep{newman2017}, though a radial gradient analysis has not been conducted. We seek here to conduct such a radial investigation and to understand if the new NFM data could potentially resolve the IMF discrepancy, or re-affirm the peculiar nature of the stellar populations in \snl1.
\subsection{Generalised Full-Spectral Fitting}
To be able to measure the IMF directly in the centre of \snl1, we employ a highly-flexible spectral-fitting code known as the Absorption Line Fitter \citep[{\sc alf};][]{conroy2018}. {\sc alf} uses the extended MILES+IRTF empirical SSP library \citep{villaume2017} to fit for stellar kinematics, ages, \(19\) individual elemental abundances, and the stellar IMF. {\sc alf} uses a \ft\ version of the {\sc emcee} package \citep{foreman-mackey2013} to sample the posterior distribution of its large parameter-space, which includes the physical parameters mentioned above, as well as a number of technical nuisance parameters to account for instrument artefacts and atmospheric corrections.\par
This kind of fitting is highly demanding on the input spectra, and so for our restricted FoV, we re-bin the data-cube to a target \(S/N=80\). Although this sacrifices some spatial resolution, it is necessary to robustly constrain the highly-flexible fits of {\sc alf}. We still resolve the new FoV with 10 spatial bins across. Given the specific aims and data quality of our observations, we run a non-standard fitting setup of \tfo{alf}. We do not fit all \(19\) elemental abundances for every spatial bin, given that even the increased \(S/N\) may not be sufficient to constrain them all simultaneously. Instead, we first construct an aperture spectrum by integrating over the circular FoV we derived earlier (still masking the dusty pixels to avoid a biased spectrum). We fit this single spectrum with the full flexibility of \tfo{alf}. The results of the fit are then used to fix many of the elemental abundances when we subsequently fit the individual binned spectra. In this way, we limit the flexibility of \tfo{alf} while retaining the ability to directly measure the IMF. Namely, we fix the elemental abundances of \nuclide{[Si/H]}, \nuclide{[K/H]}, \nuclide{[V/H]}, \nuclide{[Cr/H]}, \nuclide{[Mn/H]}, \nuclide{[Co/H]}, \nuclide{[Ni/H]}, \nuclide{[Cu/H]}, \nuclide{[Sr/H]}, \nuclide{[Ba/H]}, and \nuclide{[Eu/H]} to the best-fitting values (maximum likelihood of the posterior distributions) derived from the full fit to the aperture spectrum. We opt to keep the abundances of \chemOH, \chemCH, \chemNH, \chemNaH, \chemMgH, \chemCaH, and \chemTiH\ free, as these have a more direct impact on the measured IMF, and have been measured in some galaxies to have significant gradients \citep[e.g.][]{labarbera2019,labarbera2021,parikh2024}. The results of the aperture fit are shown in \cref{app:aperalf}.\par
In both runs of {\sc alf}, the IMF is parametrised as a two-part broken power-law, defined by the slopes \(\alpha\) (a free parameter) for \(m_\star \in [\numrange[range-phrase=\ensuremath{,{}}]{0.08}{1.0})\ \si{\Msun}\), and \(\alpha_{\rm Salpeter} = 2.3\) for \(m_\star \in [1.0,{}m_{\rm max}]\ \si{\Msun}\). We leave the default \(m_{\rm max}=\SI{100}{\Msun}\). To sample the posterior distributions, we run {\sc alf} with \(\num{1024}\) walkers. These walkers do \(\num{10000}\) burn-in steps each, a further \(\num{10000}\) burn-in after reinitialisation, then a final \(100\) steps to sample the posteriors from which the best-fit values are determined. Such a set-up is robust against local minima, and thoroughly searches the large parameter-space. This scheme is used for both the integrated-aperture and binned spectra.

\subsection{Results}
\begin{figure}
    \centerline{
        \includegraphics[width=\columnwidth]{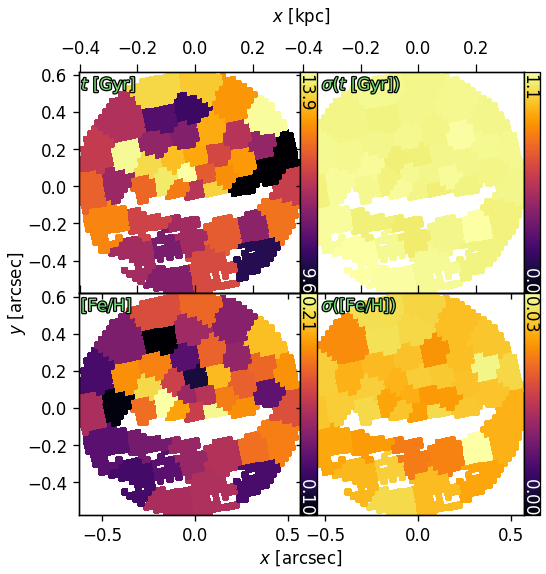}
    }
    \caption{{\em Left:} SSP-equivalent mean stellar age ({\em top}) and mean stellar metallicity \chemFeH\ ({\em bottom}) from \protect\tfo{alf}. {\em Right:} The corresponding \(1\sigma\) uncertainties derived from the posteriors of the \protect\tfo{alf} fit to each Voronoi bin.}
    \label{img:afhagemetal}
\end{figure}

We present \ND{2} maps for a number of properties of interest, generated using the maximum likelihood solution at every spatial location. The SSP-equivalent mean stellar age and mean stellar metallicity are shown in \cref{img:afhagemetal}.
These results cover a small dynamic range, and we see that the scatter between spatial bins is larger than any systematic gradient which may be present.\par
\begin{figure}
    \centerline{
        \includegraphics[width=\columnwidth]{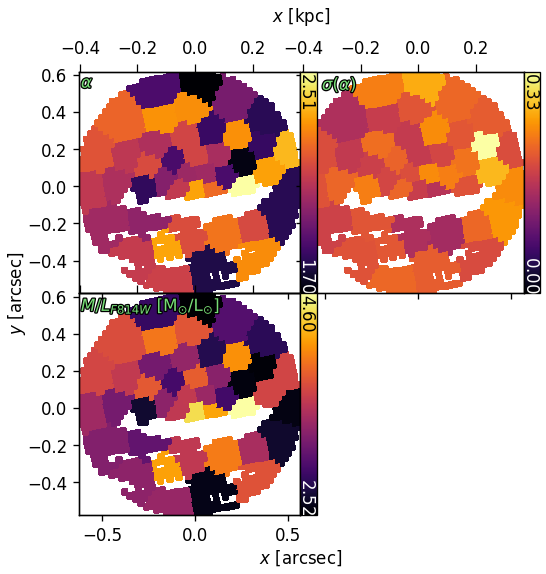}
    }
    \caption{As \protect\cref{img:afhagemetal}, but for the low-mass slope of the IMF \(\alpha\) ({\em top}) and \(F814W-\)band \mlStar\ ({\em bottom}). The \mlStar\ is derived from the outputs of \protect\tfo{alf}, rather than being fit for directly on the spectra. It therefore has no posteriors.}
    \label{img:afhimfml}
\end{figure}

\tfo{alf} also fits for the low-mass slope of the IMF directly. The top panel of \cref{img:afhimfml} shows the measured \(\alpha\), describing the mass range \(m_\star \in [\numrange[range-phrase=\ensuremath{,{}}]{0.08}{1.0})\ \si{\Msun}\).
We see that, in agreement with \cite{newman2017}, the IMF slope across the region covered by the NFM is on average consistent with a Salpeter IMF, implying a high concentration of low-mass stars. But there is little other structure in the IMF map, especially with regards to radial gradients.\par
The corresponding \(\mlStar_{F814W}\) for the measured age, metallicity, and IMF slope is shown in the bottom panel of \cref{img:afhimfml}. The \mlStar\ should closely track the variations, or lack thereof, of the IMF map in the panel above. Small deviations can be caused by additional variations in the age and metallicity.\par
\begin{figure*}
    \centerline{
        \includegraphics[width=\textwidth]{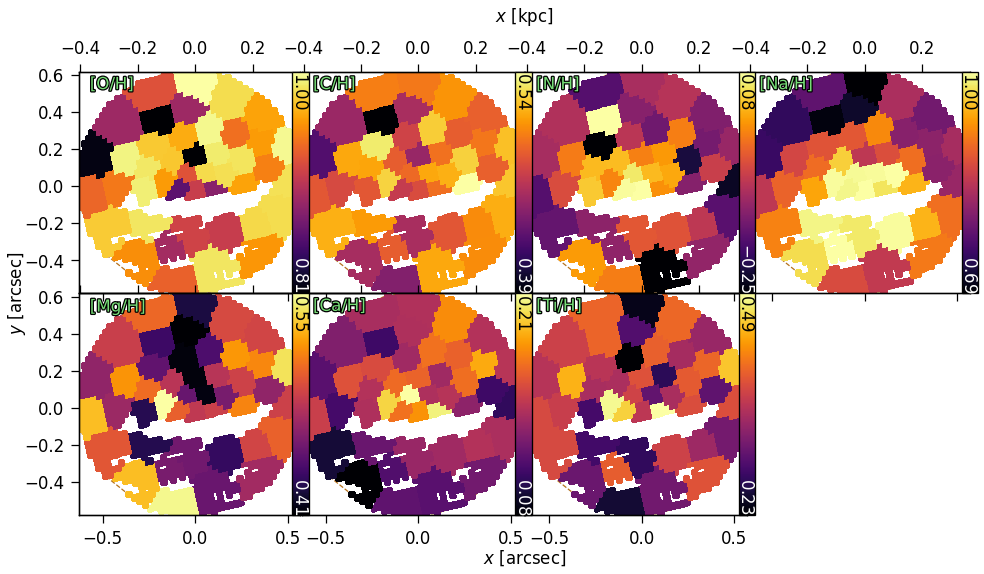}
    }
    \caption{Elemental abundances as measured in \protect\tfo{alf}.}
    \label{img:abund}
\end{figure*}

Finally, \cref{img:abund} shows maps of the remaining elemental abundances which are left free in our fits; namely, \chemOH, \chemCH, \chemNH, \chemNaH, \chemMgH, \chemCaH, and \chemTiH.
We see that there is no significant structure in any of these properties either, with the exception of the tentative evidence for gradients in \chemNaH\ and \chemNH. Moreover, the abundances of \chemOH\ and \chemNaH\ are markedly higher than the other elements.\par
\begin{figure}
    \centerline{
        \includegraphics[width=\columnwidth]{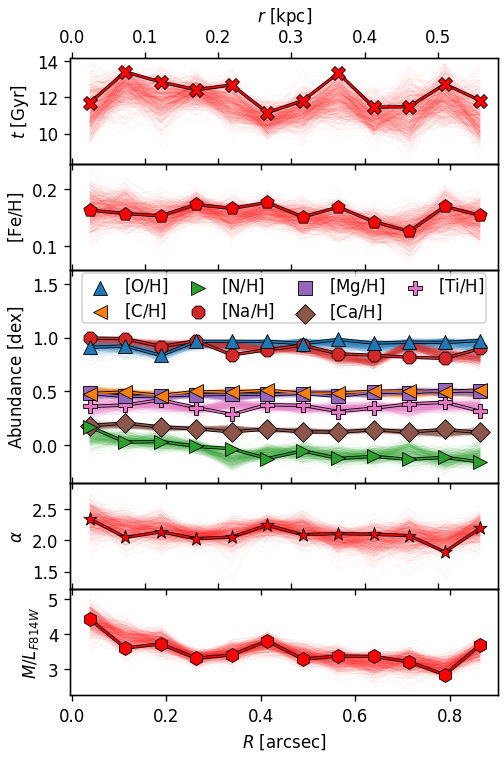}
    }
    \caption{Radial profiles of a number of stellar-population properties measured from \tfo{alf}; from top to bottom, SSP-equivalent mean stellar age, stellar metallicity \chemFeH, the individual elemental abundances which were left free to vary, the low-mass IMF slope, and the \(\ml[\star]_{F814W}\). For every measured property, we re-created \ND{2} maps by randomly sampling the posterior distributions of all spatial bins, and re-measured the radial profiles. This was repeated \(500\) times, and are shown as light curves in the background of each panel.}
    \label{img:afhradial}
\end{figure}

For a more quantitative assessment on the presence of any gradients, we show radial profiles of all of these stellar-population properties in \cref{img:afhradial}.
These profiles are computed by taking the median value within concentric elliptical annuli with the ellipticity of the stellar surface brightness MGE. Corroborating the expectation from the \ND{2} maps, we recover flat gradients in the measured properties. These data, however, probe only the central \(\sim\SI{350}{\parsec}\). This is interior to the typical transition radius of previous studies \citep[\(\SI{1}{\kilo\parsec}\); e.g.][]{labarbera2017,vandokkum2017}. It is thus unclear how prevalent gradients on the spatial scales studied here are, although they have been observed with spectral indices in M31 on scales of \(\sim\SI{100}{\kilo\parsec}\) \citep{labarbera2021}.\par
We see that some spatial bins, especially near the dust mask, exhibit strongly double-peaked posterior distributions in some parameters. In addition, we select the maximum-likelihood solution from the posterior, which can be slightly offset from the mode of the distribution for any finite number of MCMC samples. For these reasons, the randomly-sampled curves in \cref{img:afhradial} sometimes deviate from the measured curve.

\section{Ionised Gas}\label{sec:gas}
In addition to absorption-line stellar properties, we measure emission-line ionised-gas kinematics from the new data cube. Because the ionised gas emission is faint compared to the starlight, we first fit and subtract a model for the stellar continuum using {\sc pPXF}. The treatment here differs from the usage of {\sc pPXF} in \cref{sec:stelkin} for measuring stellar kinematics, since the appropriate smoothing scale for the gas will differ from that of the stars. We do not in this case need to measure reliable stellar kinematics, but rather reliably estimate the stellar continuum.\par
For every pixel in the central \(1.25\times\SI{2.50}{arcsec^2}\) region of the cube, we extract a spectrum spanning \(\qtyrange[range-phrase=-]{4680}{7150}{\angstrom}\) (rest-frame) from a circular aperture with radius varying linearly from \(\qtyrange[range-phrase=-]{0.05}{0.30}{\arcsecond}\), according to distance from the galaxy centre. We then mask the wavelength regions corresponding to the \(\nuclide{H}{\beta}\), \(\nuclide{[\atomic{O}{iii}]}5007+4959\), \(\nuclide{H}{\alpha}\), and \(\nuclide{[\atomic{N}{ii}]}6548+6584\) emission lines, regardless of whether emission is detected or not. This spectrum is then fit using simple stellar population spectral templates from \cite{vazdekis2016}, using multiplicative and additive polynomials of orders \(10\) and \(1\), respectively, to provide a high degree of flexibility. The resulting continuum model is subtracted from the single observed target pixel, thereby yielding a residual emission-line cube retaining the full spatial resolution of the original data as far as possible.\par
\begin{figure}
    \includegraphics[width=\columnwidth]{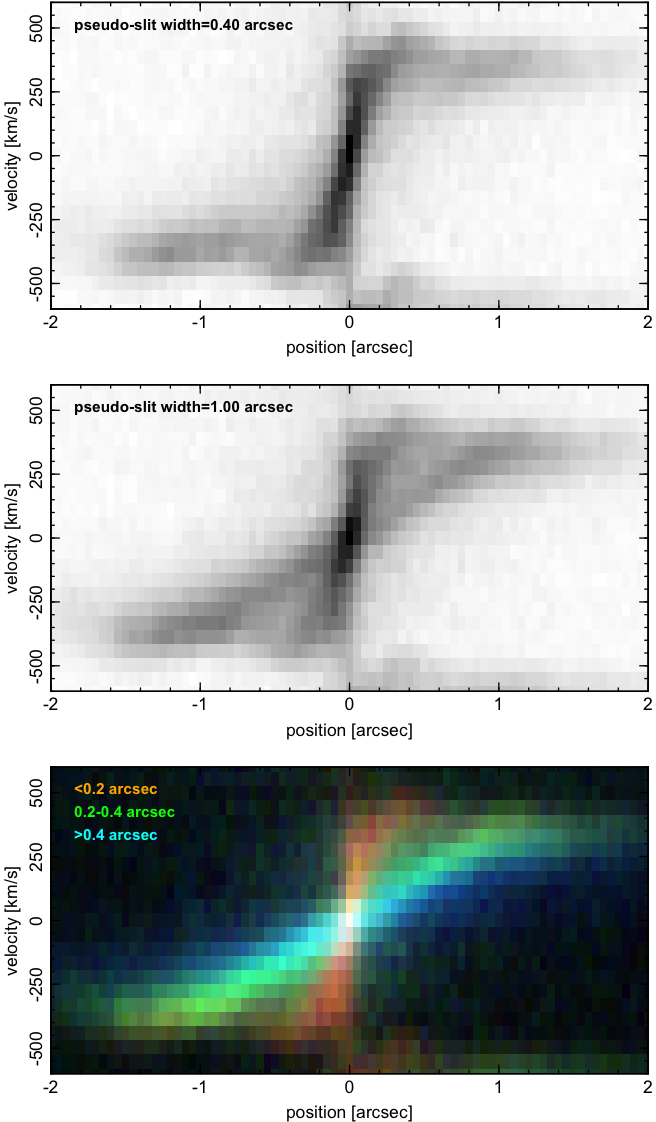}
    \caption{Major-axis position--velocity diagrams for the \([\atomic{N}{ii}]\) \(\SI{6584}{\angstrom}\) line, for two different pseudo-slit widths (upper panels) and a colour rendering (lower panel) in which the red channel reproduces the upper panel, while green and blue show pixels further from the major axis. Contamination from the \(\nuclide{H\alpha}\) line can be seen at the bottom right of each figure.}
    \label{img:gaspvd}
\end{figure}
\cref{img:gaspvd} shows position--velocity diagrams (PVDs) from the \([\atomic{N}{ii}]\) line. This fit was done by collapsing the continuum-subtracted cube onto pseudo-slits (given in the figure), then measuring the full LOSVD of the \([\atomic{N}{ii}]\) line profile at every position along the slit. This process allows for complexity in the gas kinematics, including for instance multiple co-spatial kinematic components.\par
The PVDs reveal two distinct kinematic components: a sharply rising rotation at the centre, contributing only within about \(\pm\SI{0.2}{\arcsecond}\) from the major axis, and a shallower component reaching to \(\sim\SI{2}{\arcsecond}\). The resulting X-shaped structure is of a form commonly observed in barred galaxies, where it  arises from gas trapped in orbital resonances \citep[e.g.][]{athanassoula1999}.\par
To extract a simple velocity field, we fitted the \(\nuclide{H\alpha}+[\atomic{N}{ii}]\) complex in the emission datacube with a three-Gaussian model. The three lines were constrained all to have the same LOSVD (c.f. the PVD fits above), while the \(\nuclide{H\alpha}\) and \([\atomic{N}{ii}]\) doublets have free amplitude (but the ratio between the \([\atomic{N}{ii}]\) line amplitudes is fixed). An important caveat to this is that the central regions of \snl1\ are not in fact well-described by a single LOSVD, as may be expected from the complex PVD structure. We fitted the three-Gaussian model to spectra extracted from circular apertures centred on each pixel, with a range of sizes, implementing a crude adaptive smoothing scheme.\par
The fit started with an aperture of radius \(\SI{0.400}{\arcsecond}\) (16 pixels) around the target pixel, and reduces the size by factors of two, to a final radius of \(\SI{0.025}{\arcsecond}\) (1 pixel). Each fitting step was initialised from the preceding (larger) aperture, to help avoid false minima at the smaller smoothing scales. The final velocity at each pixel is that obtained with the smallest aperture size yielding a peak signal-to-noise ratio \(S/N>3\). In practice around 50 per cent of the map pixels have an effective smoothing smaller than the nominal PSF core, and pixels with larger smoothing are at larger radius where fine spatial structure is not expected.\par
\begin{figure*}
    \includegraphics[width=\textwidth]{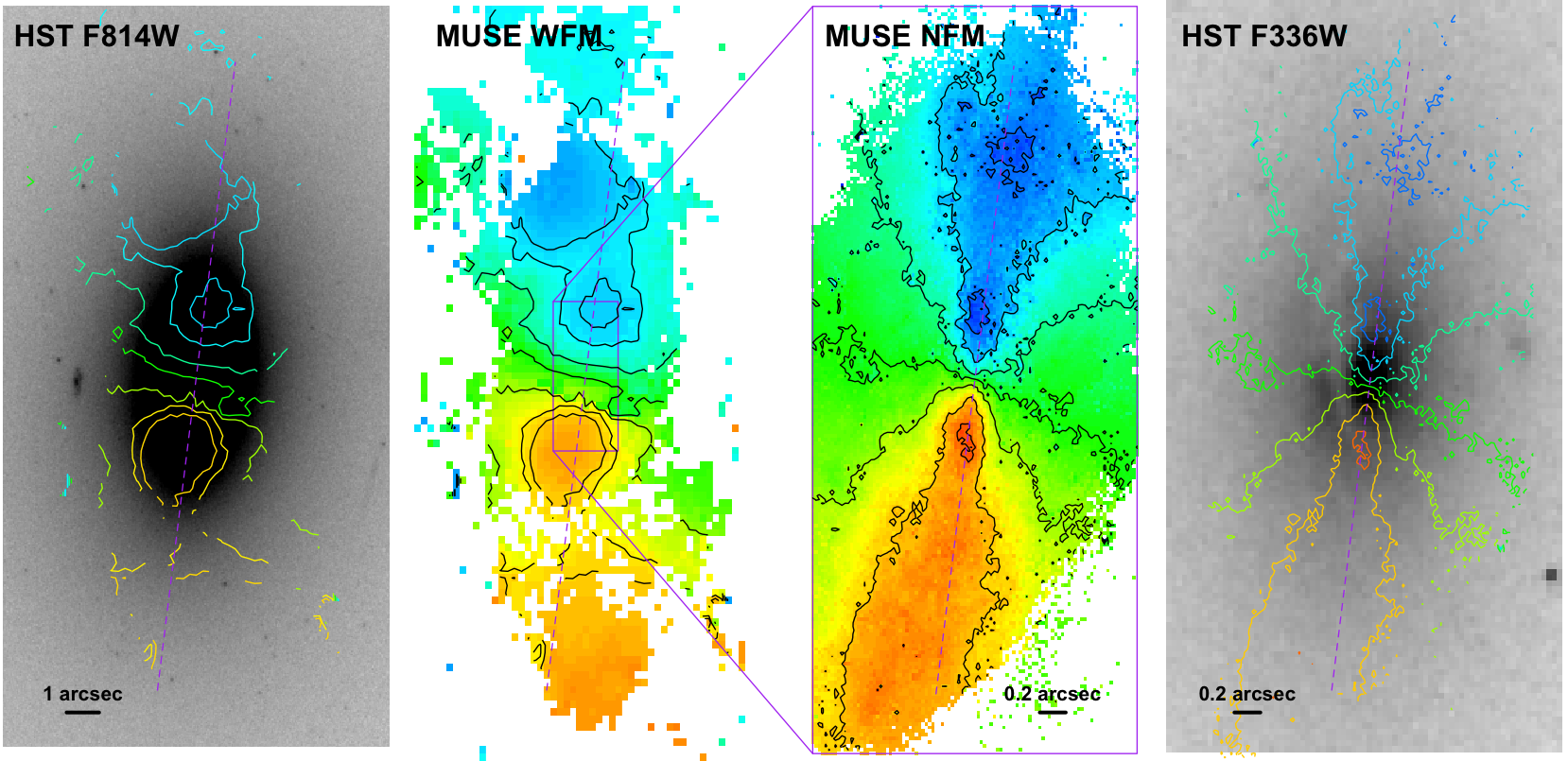}
    \caption{Gas velocities derived from the WFM and NFM observations. The left-most panel shows the WFM velocities as contours overlaid on the \(HST\) \(F814W\) image, while the colour map in the next shows more clearly the extent of the gas measurements. The third and fourth panels show the inner structures derived from the NFM data, and a comparison to the \(HST\) \(F336W\), respectively. In each pair of panels, the velocity contours are chosen to highlight relevant features; in the WFM they are drawn at \(0, \pm 150, \pm 260, \pm 290\ \si{\kilo\metre\per\second}\), showing the global rotation and outer warp, while for NFM they are at \(0, \pm 150, \pm 300, \pm 400\ \si{\kilo\metre\per\second}\), emphasising the fast-rotating inner disk. The purple dashed line in all panels is the stellar kinematic PA as a guide.}
    \label{img:gasvels}
\end{figure*}
The resulting velocity field is shown in \cref{img:gasvels}, alongside the equivalent map derived from the lower-resolution MUSE WFM data of \cite{poci2022}, with {\em HST} images from \cite{collier2018a} provided for context. On scales similar to the effective radius \((\sim\SI{3.5}{\arcsecond})\) the WFM data confirm the presence of rapid major-axis rotation in the same sense as the stellar kinematics, while beyond \(\sim\SI{5}{\arcsecond}\), the rotation continues but warps to a different position angle. The new NFM observations trace the rotation signature inwards to small scales, where it appears to peak at \(\sim\SI{0.4}{\arcsecond}\) (much larger than the nominal PSF core), corresponding to the inner structure already seen in the PVDs in \cref{img:gaspvd}. The iso-velocity contours in the NFM region show some slight asymmetry, indicating deviation from circular motion \citep[e.g.][]{spekkens2007}.\par
\begin{figure}
    \includegraphics[width=\columnwidth]{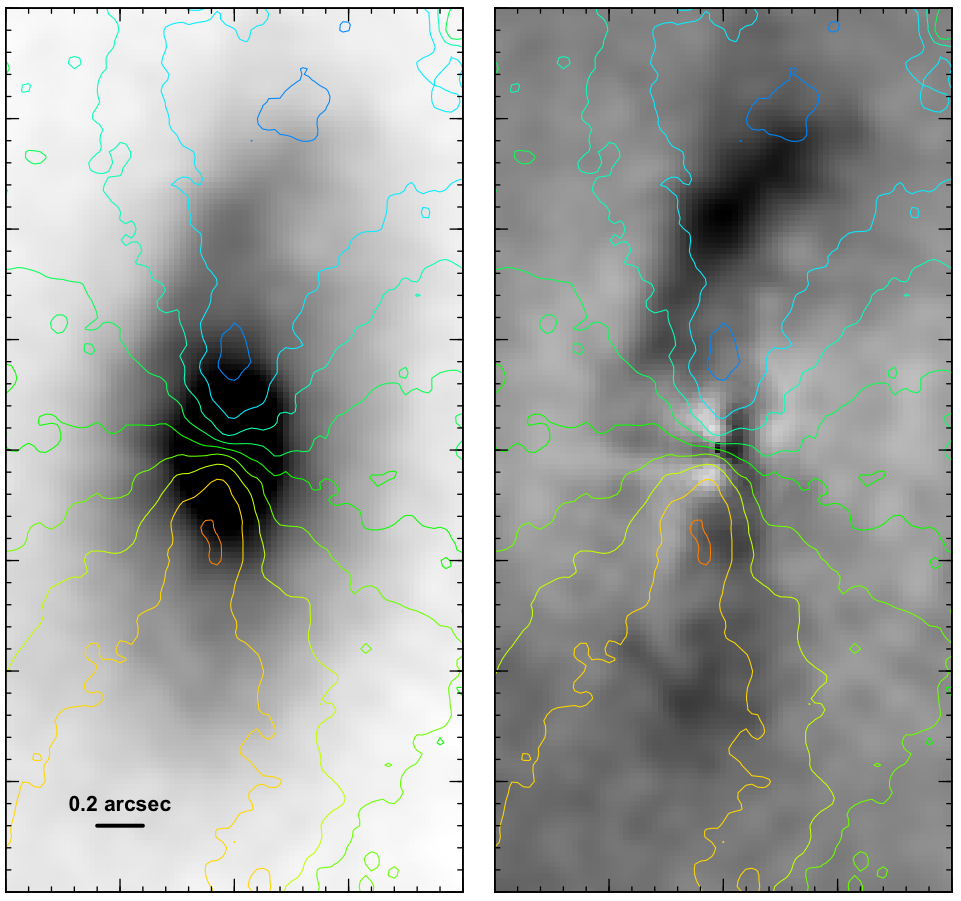}
    \caption{
    {\em Left:} Collapsed \([\atomic{N}{ii}]\) emission line image (approximately corrected for \(\nuclide{H\alpha}\) contamination), with velocity contours overlaid. Both fields have been smoothed to enhance visibility of the main structures. {\em Right:} the same after subtracting a simple elliptically-symmetric model, highlighting the spiral features.
    }\label{img:velflxoverlay}
\end{figure}
Finally, the integrated \([\atomic{N}{ii}]\) emission line image is presented in \cref{img:velflxoverlay}, and shows that the emission surface brightness rises sharply inside the \(\pm\SI{0.4}{\arcsecond}\) velocity peaks. A distinct spiral-like enhancement is clearly visible on the northern (approaching) side, at \(\qtyrange[range-phrase=-]{0.6}{1.2}{\arcsecond}\) from the galaxy centre. Such structures have been seen in ionised gas in  spiral galaxies \citep[e.g.][]{mitchell2015,kam2015,dellabruna2022,emsellem2022}, but few ETG have been studied with this spatial resolution. Together, \cref{img:gaspvd,img:gasvels,img:velflxoverlay} reveal complex structures in the inner regions of SNL-1, including features more associated with spiral and barred systems than with classical massive elliptical galaxies. 

\section{Comparing \snl1\ to the Galaxy Population}
We detect no internal variations in most of the stellar population properties of \snl1 over the small spatial region studied here. Although few comparable galaxies have been observed at this spatial resolution, it appears that other massive ETG typically harbour stronger radial gradients. In particular, past works \citep[e.g.][]{labarbera2017, conroy2017} have observed much more significant gradients in the `IMF'. However, we emphasise that there are numerous differences between these past works, as well as our own, including the spatial resolution and extent of the data, the assumed functional form of the IMF and what is measured by each `IMF' parameter --- over and above any intrinsic differences in the target galaxies of each work. Our NFM data is limited to the region which is unresolved by the typical PSF of previous works. A joint, self-consistent analysis of multi-scale data-sets will ultimately be able to elucidate any gradients in \snl1\ specifically.\par
Where comparable, our spectral results are in some cases consistent with previous works, and in other cases show significant differences. Compared\footnote{Most previous works report elemental abundances with respect to Iron, \(\nuclide{[X/Fe]}\). When comparing our results, therefore, we apply the straight-forward conversion \(\nuclide{[X/Fe]} = \nuclide{[X/H]} - \nuclide{[Fe/H]} \approx \nuclide{[X/H]} - 0.16\) (the value from the aperture spectrum fit).} to the previous work \citep{gu2022b,cheng2023,denbrok2024}, the results for \snl1\ indicate that it sits at the metal-rich end of the samples studied in those works. In fact, many elements show especially good agreement with the BCG sample of \cite{denbrok2024}, including those which we only measure on the aperture spectrum (\cref{app:aperalf}). The exceptions are oxygen and sodium, which are substantially elevated in \snl1. While modelling systematics may be playing a role here, especially in the regime of the extreme abundances measured, it is also difficult to compare disparate (and relatively small) samples of galaxies in such a detailed manner due to the possibility of substantial intrinsic scatter. In each of those past works, at most one galaxy from each sample has similar velocity dispersion to \snl1, indicating that the physical properties and/or formation histories may also differ. Thus, while the abundances of oxygen and sodium are relatively high compared to other available measurements, we can not exclude the possibility that the abundances in \snl1\ are truly high.\par
More broadly, our models must be taken in light of the particular properties of \snl1\ in the context of the galaxy population as a whole. \snl1\ exhibits some unusual characteristics, further highlighted by our new NFM data. The stars in \snl1\ simultaneously exhibit a peak velocity dispersion of \(\sim \SI{350}{\kilo\metre\per\second}\) and peak rotation velocity of \(\sim \SI{250}{\kilo\metre\per\second}\) \citep[at larger radii than shown here;][]{poci2022}. It is highly compact (\cref{tab:props}), and morphologically appears to be a quintessential massive elliptical galaxy. In contrast, galaxies in the compact ETG samples of \cite{deeley2023} and \cite{grebol-tomas2023} do not reach that magnitude in velocity or velocity dispersion, much less both simultaneously. Only one object from the catalogue of compact galaxies of \cite{spiniello2023} exceeds this velocity dispersion. Interestingly, from the parent sample of SNELLS, selected specifically to have high velocity dispersion, only 3 other galaxies reach a comparable magnitude \citep{smith2015a}. Overall, therefore, \snl1\ occupies the high-velocity-dispersion tail of the galaxy population.\par
The star-formation history\footnote{\tfo{alf} measures single-stellar population- (SSP-)equivalent properties, meaning there is no constraint on the SFH of a given spectrum. To explore the SFH, we have additionally fit the same spectra in {\sc pPXF} using the E-MILES \citep{vazdekis2016} SSP library built on the `BaSTI' isochrones \citep{pietrinferni2004} at fixed \cite{kroupa2002} IMF. As expected, the SFH is extremely short, with no weight given to models with \(t<\SI{13}{\giga\year}\). Although the underlying models and fitting methodologies differ, we see consistent constraints on the age of \snl1 between {\sc pPXF} and \tfo{alf}.} (SFH) of \snl1\ indicates that it assembled very early, and its stellar populations have remained largely unchanged since. The early and rapid SFH of \snl1\ is comparable to relic galaxies \citep{spiniello2023}; galaxies which are believed to have formed only through a single epoch of star-formation at early times with no subsequent assembly events. Given its relatively larger size compared to canonical relics, one might expect \snl1\ to have experienced a small amount of minor mergers that contributed to its size growth. It's possible that such events brought in much of the gas that is observed in the present day. While this gas resides in a region where the stellar velocity dispersion is in excess of \(\SI{300}{\kilo\metre\per\second}\), the molecular gas itself is observed to be cold \citep[\(\lesssim\SI{10}{\kilo\metre\per\second}\);][]{pociSNELLSHDII}. However, it has been proposed by low-redshift observations that even cold gas in such extreme physical conditions can be prevented from forming stars due to the shear disrupting the molecular clouds that would otherwise exist \citep{liu2021a,davis2022,lu2024a}. This may be why the SFH exhibits no signs of recent \((t<\SI{13}{\giga\year})\) star-formation activity.\par
\snl1\ also harbours compact nuclear gaseous and old stellar disks, and a prominent nuclear dust disk. The presence and structure of these disks (\cref{sec:gas}), the morphology of the dust on larger spatial scales, as well as the best-fitting \ND{3} orientation from the stellar dynamical model, provide circumstantial evidence for a stellar bar. Our dynamical modelling method does not explicitly account for bars, and the models occupy stationary gravitational potentials. However, using mock data for barred galaxies, \cite{zhu2018a} showed that the recovered physical properties of the model, including the internal orbital properties, are not significantly affected in the presence of a bar. Should a bar be present, it is likely to be at least \(\SI{10}{\giga\year}\) old, based on the age of the nuclear stellar disk \citep{desa-freitas2023a}. Such old bars have been seen previously in spiral galaxies \citep{gadotti2015, martig2021}. Should an old bar be present in \snl1, it may imply a relative dearth of violent mergers over the evolutionary history of \snl1. We note that this does not exclude the occurrence of minor mergers. \cite{martig2021} found that minor mergers do not destroy the bar, since most of the incoming stars are deposited in the outskirts of the central galaxy \citep{oser2010, karademir2019}.\par
A key motivation for the MUSE NFM data was resolving the SMBH of \snl1. Taking our best-fit value of \(M_\bullet = \bestBHLinear\ \si{\Msun}\), the expected sphere-of-influence is \(\sim \SI{73}{\parsec}\) --- which is above the spatial resolution of the NFM data. It can also be seen from \cref{img:shwcorn} that the constraint on \(M_\bullet\) is unambiguous. Although large, the measured \(M_\bullet\) is completely consistent with the \(M_\bullet-\sigma\) relation of \cite{kormendy2013}. We can therefore exclude a peculiar SMBH mass --- both unusually massive and unusually under-massive --- as a contributor to the observed discrepancy between the IMF inferences. Moreover, there is no evidence of nuclear activity such as AGN/outflows. Thus, while the central region is structurally complex, there does not appear to be anything occurring in the galaxy nucleus which could influence our interpretation of the spectral measurements.

\subsection{IMF Tension}
The contention around \snl1\ originates from the apparently contradictory results from mass-based and spectroscopic analyses. Measurements of the mass (and mass-to-light ratios) are consistent with a \cite{kroupa2002} IMF, while spectral analyses simultaneously indicate a relative over-abundance of low-mass stars. In this work, we find similar results. Our dynamical analysis provides one measure of the \(\mlStar_{F814W}\), since the mass model is derived from the flux in \(F814W\) band. The dynamical model is consistent with \(\mlStar_{F814W} \sim 2.3\). Conversely, our spectral fits are consistent with \(\mlStar_{F814W} \sim 3.5\). At face value, this implies a dynamical mis-match parameter \(\alpha_{\rm IMF} \sim 0.65\). Moreover, the spectral analysis finds that the central region of \snl1\ is consistent with a low-mass IMF slope \(\alpha \approx \alpha_{\rm Salpeter}\). This implies an abundance of dwarf stars significantly higher than a MW-like IMF.\par
A possible resolution (though not the only one) to the above tension proposed previously \citep{smith2014} suggests that the two techniques probe different physical scales of the galaxy. In that scenario, the spectral analyses are biased towards the centre of the galaxy, owing to the fact that the (bright) central stellar populations dominate the spectral signatures. Conversely, the lensing/dynamical analyses provide a more global mass estimate, which would contain greater contributions from the outer, supposedly less-extreme, stellar populations. We are able to exclude this as a possible resolution to the tension, as our spectral and dynamical analyses are carried out over the same FoV. We have seen that spatial gradients in these properties over this FoV are minimal, and yet the discrepancy persists. This is seen similarly in \cite{lu2023a}.\par
The different measurement techniques are, however, sensitive to different regions of the IMF \citep[e.g.][]{lu2023}. In the case of \tfo{alf} as used in this work, the `IMF' parameter is probing the mass range \(m_\star \in [0.08, 1.0)\ \si{\Msun}\). Alternatively, for example, the free parameter of the \cite{vazdekis1996} IMF, and its subsequent incarnations \citep[e.g.][]{martin-navarro2019}, constrains the mass range \(m_\star \in [0.4, 72)\ \si{\Msun}\). \cite{smith2020a} describes in detail the variety of assumptions intrinsic to the different spectral-fitting approaches. Yet in contrast to spectral analyses, dynamical/mass-based measurements are naturally sensitive to the full mass-range of the IMF, given that they probe the integral over every episode of star formation in a galaxy's history. As such, we might not expect these measurements to agree even in principle. For example, when compared to measurements from \tfo{alf}, the mass (and mass-to-light ratio) methods might disagree for any value of \(\alpha_{\rm high} \neq \alpha_{\rm Salpeter}\), where \(\alpha_{\rm high}\) is the unknown true slope of the IMF over \(m_\star \in [1, 100]\ \si{\Msun}\). The fitting approach in \tfo{alf} is justifiably driven by the lack of observational constraints on the high-mass end of the IMF for (even mildly) old integrated stellar spectra, but it means that comparisons to other techniques must be more cautious. Different mass regions of the IMF could be sensitive to, among other things, multiple epochs of star formation \citep{denbrok2024}, which would be captured by mass-based measurements but not spectral analyses. Direct constraints on the high mass end of the IMF can be obtained using resolved studies of actively star-forming regions \citep[e.g.][]{yan2022b}, notwithstanding the complication of extremely short lifetimes of the most massive stars \citep{wirth2022}. But even in this case, such samples would be restricted to the small range of physical conditions found in the Solar neighbourhood, and unable to explore the expected dependence of the IMF on, for instance, metallicity \citep[e.g.][]{martin-navarro2015}, density \citep[e.g.][]{poci2022a}, and natal conditions \citep{cameron2023}. More importantly, it is unknown whether those results could be translated to the conditions present in ETG, in any case.\par
It is possible, then, that the reported tension in the literature is dramatically weaker or entirely resolved simply by the fact that the different measurement techniques are sensitive to different mass regimes of the IMF. Interestingly, \cite{smith2020a} found that if the differences in the mass-to-light ratios are to be accounted for solely through the differences in the methodolgies, the low-mass cut-off mass of the IMF, \(m_{\rm cut}\), would need to be \(\sim\SI{0.15}{\Msun}\). This is because increasing the \(m_{\rm cut}\) would significantly lower the spectroscopic \mlStar, without affecting the spectral fit (since such low-mass stars don't contribute any light to the spectrum anyway).\par
Unfortunately, empirically constraining \(m_{\rm cut}\) is difficult. On one hand, a high abundance of stars below this mass have been detected within the MW \citep[e.g.][]{sollima2019}, implying that they readily form. Similarly, the spectral fits of \cite{conroy2017} seem to require a monotonic increase in the number of stars inversely proportional to their mass, all the way down to the Hydrogen-burning limit of \(m_\star = \SI{0.08}{\Msun}\). \cite{chabrier2014} find that in extreme densities and turbulent gas, the peak mass of the IMF shift towards lower mass with respect to spiral galaxy (e.g. MW)-like conditions. They find that these conditions are conducive to IMF more dwarf-rich than even that of \cite{salpeter1955}. On the other hand, however, the simulations of \cite{bate2023} show that at high redshift, there could be significant variation in \(m_{\rm cut}\) as a function of metallicity, such that star-formation at Solar metallicity at \(z=5\) produces a mass function which is deficient in stars with \(m_\star \lesssim \SI{0.10}{\Msun}\). Having Solar metallicity by \(z=5\) implies a chemical enrichment history which was very rapid, but one which is compatible with the data of \snl1. Elevated \(m_{\rm cut}\) has also been inferred from observations in \cite{barnabe2013}, by anchoring the expectation from the spectral fitting to that from lensing and stellar dynamics. There, circumstantial evidence for \(m_{\rm cut} \approx \qtyrange[range-phrase=-]{0.1}{0.15}{\Msun}\) was found for two massive ETG using high resolution aperture spectra. \cite{newman2017} also tested for the variation of \(m_{\rm cut}\) by including it as a free parameter in their spectral fits, and find a peak in the posterior at \(m_{\rm cut} = \SI{0.15}{\Msun}\) for \snl1. We re-did the spatially-integrated  spectral fit discussed in \cref{app:aperalf}, but this time allowing \(m_{\rm cut}\) to vary. This fit has a best-fitting \(m_{\rm cut}\approx 0.35\spm{0.0064}{0.19}\ \si{\Msun}\). Despite the high \(S/N\) of this integrated spectrum, the posterior distribution of \(m_{\rm cut}\) is still broad. We conclude that \(m_{\rm cut}\) is unconstrained by the data, especially for the spatially-resolved spectra. Consequently, resolving the apparent tension in the literature will require higher-quality data and improved models. Nevertheless, if the scenario described in \cite{smith2020a} is applicable to \snl1\ and its IMF had a relatively high \(m_{\rm cut}\), differences in the modelled mass ranges of the IMF may partially resolve the observed discrepancy in the \mlStar. Whether this can fully explain the observations, however, remains to be seen.

\section{Conclusion}
We have analysed new VLT/MUSE NFM data of a nearby strong-lens, \snl1. We have applied flexible spectral and dynamical methods to disentangle the contributions to the mass budget in the nuclear region. In particular, we have measured the IMF with high spatial resolution, and sought to resolve an existing tension between different measurement techniques. Our results are summarised here:
\begin{itemize}
    \item Dynamically, we modelled a previously-undetected nuclear stellar disk. We find it be highly flattened, embedded in a largely elliptical mass distribution. The new data allow us to directly constrain the mass of the central SMBH. We measure \(M_\bullet = \bestBHLinear\ \si{\Msun}\).
    \item Spectroscopically, we have measured a variety of stellar-population properties. \snl1\ is extremely old, and relatively metal-rich. It is enhanced in many of the individual chemical abundances, especially \chemNaH\ and \chemOH, but with very little spatial variation over the small central region probed here. Under the assumptions of \tfo{alf} regarding the shape of the IMF and its mass limits, we have measured relatively dwarf-rich populations across the entire nuclear region, with a IMF slope of \(\alpha \approx \alpha_{\rm Salpeter}\), for the mass range \(m_\star \in [0.08, 1.0)\ \si{\Msun}\).
    \item At face value, we have recovered the existing tension in the literature regarding the IMF as measured by different methodologies. Peculiarly, the spectra predict over-massive populations with respect to the dynamics. Several causes suggested previously are excluded, including differences in the physical scales probed by each method, as well as unresolved non-stellar contributions (e.g. an unaccounted-for SMBH). With the flexibility of the models applied here, we can also rule out the presence of some biases in earlier works, such as assumptions on the orbital isotropy. Yet we conclude that a ``tension'' itself may not be physically present, and rather caution the comparison of disparate measurement techniques, especially when it concerns the IMF.
\end{itemize}
In general, future observations at this spatial scale will prove critical to understanding past IMF studies, and indeed the central regions of ETG more broadly. In a forthcoming work, we will jointly model multi-scale, multi-tracer data for \snl1, in order to confirm the nuclear structure inferred in this work, and its implications for the IMF.
Ultimately, the conundrum of the IMF in \snl1\ remains. This tension can not be decoupled from its intrinsic physical properties and complicated nuclear structures, which are also yet to be fully characterised. Multi-scale, multi-tracer data may help in breaking any degeneracies, in particular of the existing mass-based models. However, the evidence indicates that something intrinsic, at least for \snl1, is favouring these contradictory results. The complex structures present in the core of \snl1\ indicates that it may not be the ideal candidate for benchmark studies of various IMF techniques. Thus, the task remains to understand the intricate spectral signatures and dynamical structures in order to robustly measure the IMF in external galaxies.

\section*{Acknowledgements}
We thank Tim Davis, Dimitri Gadotti, and Chiara Spiniello for many insightful discussions. AP acknowledges support from the Hintze Family Charitable Foundation. AP and RJS acknowledge support from the Science and Technology Facilities Council through the Durham Astronomy Consolidated Grants 2023–2026 (ST/T000244/1 and ST/X001075/1). This research is based on observations collected at the European Organisation for Astronomical Research in the Southern Hemisphere under ESO programme 109.22X3.001. This work used the DiRAC\@Durham facility managed by the Institute for Computational Cosmology on behalf of the STFC DiRAC HPC Facility (\url{www.dirac.ac.uk}). The equipment was funded by BEIS capital funding via STFC capital grants ST/K00042X/1, ST/P002293/1, ST/R002371/1 and ST/S002502/1, Durham University and STFC operations grant ST/R000832/1. DiRAC is part of the National e-Infrastructure. This work utilised existing software packages for data analysis and presentation, including \tso{AstroPy} \citep{astropycollaboration2013}, \tso{Cython} \citep{behnel2011}, \tso{IPython} \citep{perez2007}, \tso{matplotlib} \citep{hunter2007}, \tso{NumPy} \citep{harris2020a}, the \tso{SciPy} ecosystem \citep{virtanen2020}, \tso{statsmodels} \citep{seabold2010}, and \tso{seaborn} \citep{waskom2021}. We thank the anonymous referee for their constructive comments which improved the quality and clarity of this work.

\section*{Data Availability}
The observational data used in this work are publicly available in the ESO archive (\url{archive.eso.org}). Other products can be provided upon reasonable request to the author.



\bibliographystyle{mnras}
\bibliography{{SNELLSHDi_subm.arXiv}}



%
\appendix
\section{Sky CTI Issues}\label{app:sky}
The science spectra from the pipeline-reduced data-cube exhibited strange sky emission-line shapes, characterised by a strong over-subtraction and strong under-subtraction on neighbouring spectral pixels around the emission line wavelength. To investigate the cause of this issue, we set about analysing the input data at each stage of the reduction pipeline. From this, it became apparent that the dedicated offset sky frames were problematic, which we believe is caused by charge-transfer inefficiency (CTI) effects. We describe our process and conclusions in this section.\par
The MUSE data reduction pipeline \citep{weilbacher2020} includes a sophisticated sky modelling treatment, which distinguishes between line and continuum components. For observations of targets which fill the MUSE field of view, as in our case, this sky model is derived from a blank sky field observed close in time to the science exposure. The model includes a description of the line spread function (LSF) and its variation with position and wavelength. This LSF is used, in combination with the modelled fluxes of known emission lines, to build the line component of the sky, which varies on rapid timescales. The fluxes of different groups of emission lines vary in lockstep, so that the amplitudes of the groups can be adjusted when transferring the line model to the science data. The continuum component is assumed to vary only on longer timescales, and is subtracted from the science data without modification.\par
In general the blank sky exposure is much shorter than the science observations, since it is assumed that the sky information is pooled from all \(\sim 10^5\) spatial pixels. For each of the six OB of our observation, the sky exposures were \(\SI{120}{\second}\), compared to \(\qtyrange[range-phrase=-]{2257}{2574}{\second}\) for the science target (depending on the overheads). With the \(25\ \si{mas}\) pixel size of NFM observations in particular, the short-exposure sky frames lead to extremely low count levels. Indeed, in our \(\SI{120}{\second}\) exposures, the (clipped) mean pixel value is \(0.27\ \si{e^-}\) above the bias level, i.e. most pixels receive no photons during the exposure. Under such conditions, CCD images exhibit artefacts caused by charge being temporarily retained by empty ``traps'' in the silicon lattice during read-out. One observable consequence of such CTI effects is the formation of near-exponential trails behind sources, extended away from the readout direction, caused by the late arrival of temporarily trapped charge \citep[e.g. see][]{massey2010}.  By contrast in much longer exposures (e.g. the science frames), or for MUSE WFM observations with \(64\times\) larger pixel area, most charge traps will be filled during the exposure, and become unable to retain electrons during the read-out phase.\par
\begin{figure*}
    \centerline{\includegraphics[width=\textwidth]{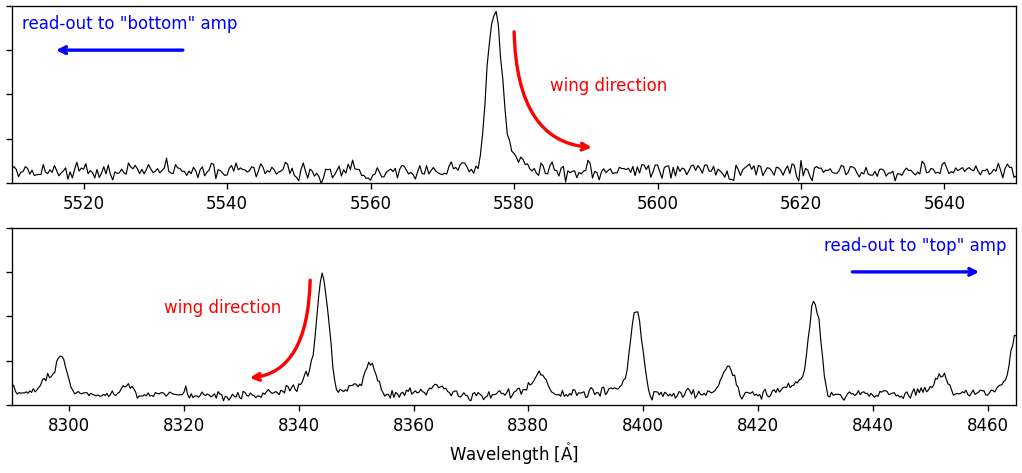}}
    \caption{Profiles of sky emission lines, binned over large areas of the short offset-sky cubes, showing CTI-induced wings extending away from the amplifier. }\label{img:skywings}
\end{figure*}

In the case of our MUSE NFM blank sky observations, trails likely due to CTI effects are clearly observable in the observed sky line profiles, as seen in the mean calibrated spectrum (see \cref{img:skywings}). Sky lines in the red part of the spectrum fall in the upper quadrants of the CCD, and are read out to amplifiers at the top; these lines show an asymmetric wing towards the blue. By contrast, the \(\SI{5577}{\angstrom}\) line falls in the lower quadrants, which are read out to amplifiers at the bottom, and the line profile asymmetry is reversed as expected.\par
The asymmetric line profiles are observed only in the short blank-sky observations, and not in the sky lines in the science data. Since the pipeline assumes the LSF is identical between the sky and target observations, the ``line'' component cannot respond to the presence of the wings. Instead, they are absorbed into the sky ``continuum'' component, which then includes highly asymmetric features slightly offset from each sky line, which we believe causes the sharp over-/under-subtraction.\par
We tested many avenues to account for the CTI issues in the sky frames before including them in the data reduction process. For example, we attempted to include a correction to the sky lines which depended on the read-out direction of each line. We also tried to smooth the continuum component of the sky model, which should be where the asymmetry from the sky emission lines is absorbed, before folding it back into the data reduction. All of these attempts resulted in significant systematic noise being introduced to the science frames because of the low signal present in the sky frames to begin with.\par
In the end, we constructed a spectrum from a small annulus at the exterior of the raw science cube (\(\sim\SI{3.5}{\arcsec}/\SI{2.26}{\kilo\parsec}\) radius from the galaxy centre), which contains sky and some signal from the science target. We then subtracted that spectrum from the full raw science cube to obtain our ad-hoc `sky-subtracted' cube. The motivation for this is that the sky should have a constant shape and surface brightness across the FoV, while the steep decline in surface brightness of the target means that its signal is relatively unaffected within our science aperture of \(\SI{0.6}{\arcsecond}\) radius.

\section{Data-Cube Errors}
\begin{figure*}
    \centerline{\includegraphics[width=\textwidth]{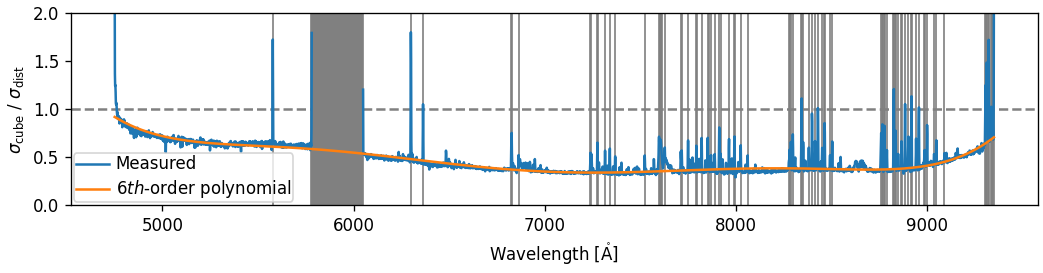}}
    \caption{Ratio spectrum of the mean reported uncertainty across a small spatial patch \(\sigma_{\rm cube}\) to the estimated true observational uncertainty over that same region \(\sigma_{\rm dist}\), shown in blue. A \(6th\)-order polynomial fit to the continuum of the ratio spectrum is shown in orange. The emission-line masking is shown in grey. The polynomial fit is used to scale the data-cube uncertainties.}
    \label{img:poly}
\end{figure*}

During the data analysis, we noted that the error cube provided by the standard ESO pipeline appeared to be significantly underestimated, based on the noisy appearance of spectra with an apparently high formal \(S/N\). This has also been seen in other data-sets \citep[see][for a discussion]{emsellem2022}. We endeavoured to quantify and rectify this for our data-cube. To this end, we first isolated a number of small (\(20\times 20\) pixels) spatial patches in the outskirts of \snl1 (outside the eventual science aperture), within which we assume that the intrinsic galactic variations are negligible. We computed the distribution of fluxes using every spaxel within a given spatial patch, for every wavelength channel. Taking the standard deviation of each distribution, \(\sigma_{\rm dist}\), results in a spectrum of the true observational uncertainty. We repeated this for every spatial patch to check for large-scale variations, finding that the error spectra were consistent across the different spatial patches. Comparing \(\sigma_{\rm dist}\) to the mean reported uncertainty from the data-cube over the same spatial region, \(\sigma_{\rm cube}\), provides an estimate for how much the latter are underestimated. We fit a general \(6th\)-order polynomial to the spectrum of the ratio \(\sigma_{\rm cube}/\sigma_{\rm dist}\) (masking the sky lines). We then scale the data-cube uncertainties by this polynomial to arrive at an uncertainty-corrected data-cube which we use throughout this work. The uncertainty spectrum and polynomial fit are shown in \cref{img:poly}.
We find that the ratio is typically \(\sim 0.5\) across the spectral range, implying that the formal uncertainties are underestimated by a factor of \(2\). This is larger than what was found in \cite{emsellem2022}, who used a different approach to estimate this level and found a \(\sim 10-30\%\) underestimation. 

\section{Schwarzschild Model Parameter-Space Exploration}\label{app:shw}
\begin{figure*}
    \centerline{\includegraphics[width=\textwidth]{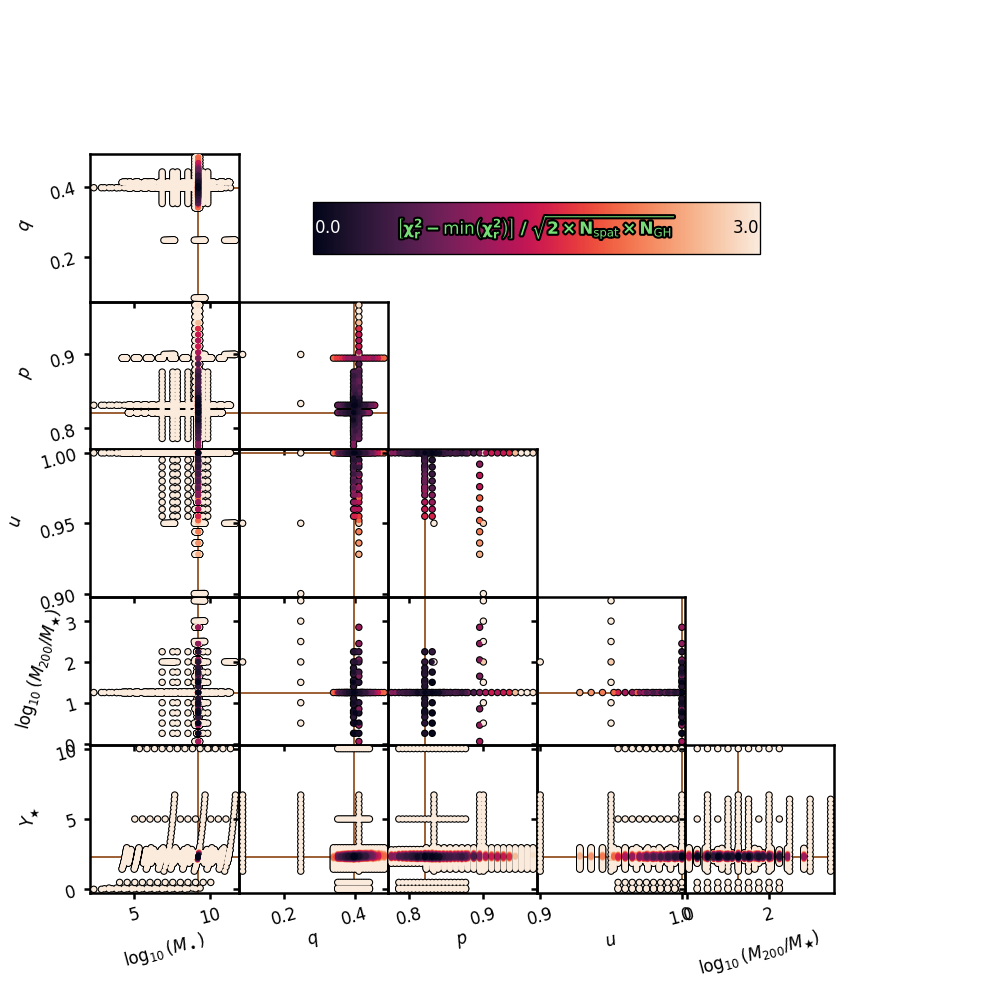}}
    \caption{As \protect\cref{img:shwcorn}, but showing the full range of parameters explored.}
    \label{img:shwcornClip}
\end{figure*}

In \cref{img:shwcornClip}, we show the full extent of the physical parameters explored for the \shw\ model in searching for the best-fit.

\section{Full \tfo{alf} Fit to Aperture Spectrum}\label{app:aperalf}
\tfo{alf} employs a multi-dimensional fitting routine in order to be able to measure an array of chemical abundances and other stellar properties. As a consequence, a fit including all the flexibility in \tfo{alf} requires input spectra with very high \(S/N\); \(100\) or greater \citep[e.g.][]{cheng2023}. As a combination of the CTI issues discussed above, coupled with the small pixel size of the NFM, our data do not meet this \(S/N\) even with significant spatial binning. By instead limiting the flexibility of the \tfo{alf} fit, we prevent broad variations of some elemental abundances which in practise would be largely unconstrained. Unconstrained variations in those abundances could trade-off with other parameters of the fit which we are directly interested in, especially in the presence of noise. Thus, removing the ability of those abundances to vary in principle produces more stable solutions for the parameters we are interested in.\par
\begin{figure*}
    \centerline{\includegraphics[width=\textwidth]{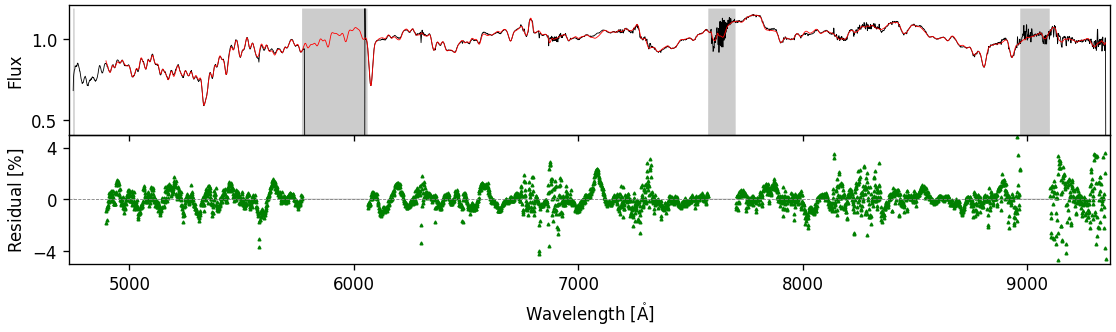}}
    \caption{{\em Top:} Spatially-integrated spectrum of \snl1\ (black), and the \tfo{alf} fit (red). {\em Bottom:} Normalised residuals of the fit [\(100\times\)(data\(-\)model)/data].}
    \label{img:apspec}
\end{figure*}
\begin{figure*}
    \centerline{\includegraphics[width=\textwidth]{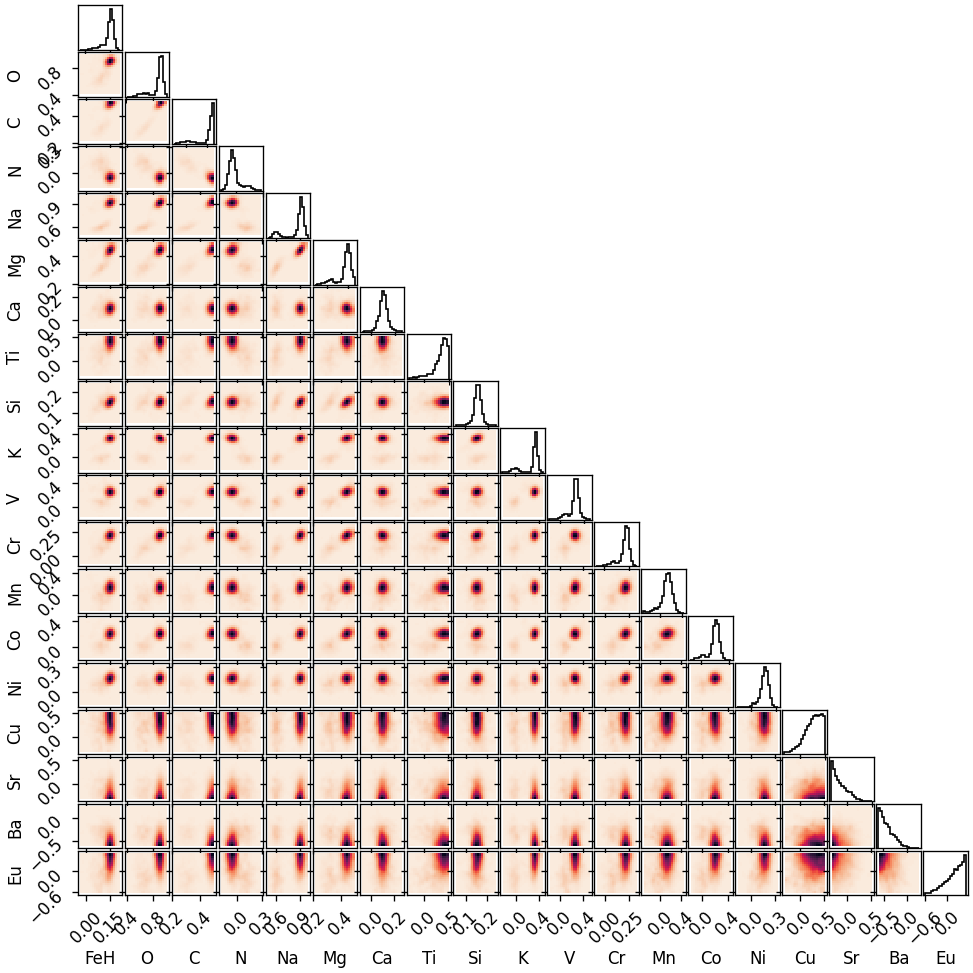}}
    \caption{Posterior distributions from the fully-flexible fit to the \SI{1.25}{\arcsecond} integrated aperture spectrum. All elemental abundances in \tfo{alf} are shown. The cross-panels are density maps of the underlying MCMC chain for clarity, with dark regions corresponding to higher likelihoods. This figure illustrates that there are no correlations between the abundances we are interested in scientifically and those we keep fixed in subsequent fits.}
    \label{img:cornFree}
\end{figure*}

We first constructed a single spectrum by integrating over the entire FoV (that which is shown in \cref{img:schw}), resulting in a spectrum with \(S/N=550\). This is then fit with the full flexibility of \tfo{alf}, including all individual elemental abundances, jitter and instrumental terms, and kinematic moments \(V\), \(\sigma\), \(h3\), and \(h4\). We keep the same parametrisation of the IMF, being a two-part power-law over the mass range \(m_\star \in [0.08, 100]\ \si{\Msun}\), with free lower-mass slope \(\alpha_1\) over \(m_\star \in [0.08, 1.0)\ \si{\Msun}\), and fixed \(\alpha_{\rm Salpeter} = 2.3\) for \(m_\star \geq 1.0\ \si{\Msun}\). This fit is shown in \cref{img:apspec}. We show a subset of the posteriors in \cref{img:cornFree} for clarity, to assess the correlations between the elemental abundances.\par
We adopt the values of \nuclide{[Si/H]}, \nuclide{[K/H]}, \nuclide{[V/H]}, \nuclide{[Cr/H]}, \nuclide{[Mn/H]}, \nuclide{[Co/H]}, \nuclide{[Ni/H]}, \nuclide{[Cu/H]}, \nuclide{[Sr/H]}, \nuclide{[Ba/H]}, and \nuclide{[Eu/H]} from this fit for all subsequent fits to the individual spectra binned to \(S/N=80\). The stellar properties of the integrated fit, including the values to which the listed elemental abundances are fixed, are given in \cref{tab:abund}.
\begin{table}
    \centerline{
    \begin{tabular}{r|S[table-format=3.2]|S[table-format=3.2]}
        {\bf Parameter} & {\bf Value} & {\(1\sigma\)}\\\hline\hline
        Age \(\log_{10}(t\ [\si{\giga\year}])\) & 1.02 & 0.03\\
        Metallicity \chemZH & 0.06 & 0.03\\
        IMF slope \(\alpha_1\) & 2.28 & 0.19\\
        Iron \chemFeH & 0.16 & 0.03\\
        Oxygen \chemOH & 0.95 & 0.10\\
        Carbon \chemCH  & 0.49 & 0.07\\
        Nitrogen \chemNH & 0.00 & 0.07\\
        Sodium \chemNaH & 0.93 & 0.12\\
        Magnesium \chemMgH & 0.47 & 0.05\\
        Calcium \chemCaH & 0.16 & 0.02\\
        Titanium \chemTiH & 0.35 & 0.13\\\hline
        Silicon \nuclide{[Si/H]} & 0.10 & 0.04\\
        Potassium \nuclide{[K/H]} & 0.35 & 0.14\\
        Vanadium \nuclide{[V/H]} & 0.36 & 0.08\\
        Chromium \nuclide{[Cr/H]} & 0.21 & 0.06\\
        Manganese \nuclide{[Mn/H]} & 0.11 & 0.11\\
        Cobalt \nuclide{[Co/H]} & 0.25 & 0.09\\
        Nickel \nuclide{[Ni/H]} & 0.18 & 0.05\\
        Copper \nuclide{[Cu/H]} & 0.48 & 0.17\\
        Strontium \nuclide{[Sr/H]} & -0.28 & 0.14\\
        Barium \nuclide{[Ba/H]} & -0.57 & 0.14\\
        Europium \nuclide{[Eu/H]} & 0.37 & 0.25\\
    \end{tabular}
    }
    \caption{Best-fitting stellar parameters from the \SI{1.25}{\arcsecond} integrated aperture spectrum and the associated \(1\sigma\) uncertainty derived from the posterior distributions. The elemental abundances below the middle rule are held fixed in subsequent spectral fits to the values listed here.}
    \label{tab:abund}
\end{table}
The implicit assumption in this approach is that there are no (strong) spatial gradients of those particular elemental abundances, but in practise we only care that they do not introduce a systematic bias in any of the parameters we are actually interested in, such as the IMF slope.

\bsp	
\label{lastpage}
\end{document}